\documentclass{agujournal2019}
\usepackage{url} 
\usepackage{soul}
\usepackage{amsmath}
\usepackage{amssymb}
\usepackage{amsmath,amsthm,mathrsfs,mathtools,amscd}
\usepackage{graphicx}
\usepackage{wrapfig}
\usepackage{lscape}
\usepackage{rotating}
\usepackage{epstopdf}

\draftfalse

\journalname{Journal of Geophysical Research - Space Physics}

\begin{document}

%
%


\title{Chromospheric Heating by MHD Waves and Instabilities}

%
%




\authors{A.K.~Srivastava\affil{1}, J. L.~Ballester\affil{2,3}, P.S.~Cally\affil{4}, M. Carlsson\affil{5,6}, M.~Goossens\affil{7}, D.B.~Jess\affil{8}, E.~Khomenko\affil{9,10}, M.~Mathioudakis\affil{8}, K.~Murawski\affil{11}, T.V.~Zaqarashvili\affil{12,13,14}}
\affiliation{1}{Department of Physics, Indian Institute of Technology (BHU), Varanasi-221005, India}
\affiliation{2}{Departament de Física, Universitat de les Illes Balears, E-07122 Palma de Mallorca, Spain}
\affiliation{3}{Institut d$'$Aplicacions Computacionals de Codi Comunitari (IAC3), Universitat de les Illes Balears, E-07122 Palma de Mallorca, Spain}
\affiliation{4}{School of Mathematics, Monash University, Clayton, Victoria 3800, Australia}
\affiliation{5}{Institute of Theoretical Astrophysics, University of Oslo, P.O. Box 1029 Blindern, N-0315 Oslo, Norway}
\affiliation{6}{Rosseland Center for Solar Physics, University of Oslo, P.O. Box 1029 Blindern, N-0315 Oslo, Norway}
\affiliation{7}{Centre for mathematical Plasma Astrophysics (CmPA), KU Leuven, Celestijnenlaan 200B bus 2400, B-3001 Leuven, Belgium}
\affiliation{8}{Astrophysics Research Centre, School of Mathematics and Physics, Queen's University Belfast, Belfast, BT7 1NN, UK}
\affiliation{9}{Instituto de Astrofísica de Canarias, E-38205 La Laguna, Tenerife, Spain}
\affiliation{10}{Departamento de Astrofísica, Universidad de La Laguna, E-38205, La Laguna, Tenerife, Spain}
\affiliation{11}{Institute of Physics, University of M. Curie-Sk{\l}odowska, 
Pl. M. Curie-Sk{\l}odowskiej 1, 20-031 Lublin, Poland}
\affiliation{12}{IGAM, Institute für Physik, University of Graz, Universitätsplatz 5, A-8010 Graz, Austria}
\affiliation{13}{Ilia State University, Cholokashvili ave 5/3, Tbilisi, Georgia}
\affiliation{14}{Abastumani Astrophysical Observatory, Mount Kanobili, Abastumani, Georgia}




\correspondingauthor{A.K.~Srivastava}{asrivastava.app@iitbhu.ac.in}




\begin{keypoints}
\item magnetohydrodynamics (MHD)
\item waves
\item instabilities
\item solar atmosphere
\item chromosphere
\end{keypoints}

\newcommand{\bv}{Brunt-V\"ais\"al\"a}
\newcommand{\rd}{\mathrm{d}}
\newcommand{\rD}{\mathrm{D}}
\newcommand{\pderiv}[2]{\frac{\partial#1}{\partial#2}}
\newcommand{\pderivd}[2]{\frac{\partial^2#1}{\partial#2^2}}
\newcommand{\pderivt}[2]{\frac{\partial^3#1}{\partial#2^3}}
\newcommand{\pderivq}[2]{\frac{\partial^4#1}{\partial#2^4}}
\newcommand{\deriv}[2]{\frac{\rd#1}{\rd#2}}
\newcommand{\derivd}[2]{\frac{\rd^2#1}{\rd#2^2}}
\newcommand{\Deriv}[2]{\frac{\rD#1}{\rD#2}}
\newcommand{\Derivd}[2]{\frac{\rD^2#1}{\rD#2^2}}
\newcommand{\mpderiv}[3]{\frac{\partial^2#1}{\partial#2\partial#3}}
\newcommand{\vdot}{{\boldsymbol{\cdot}}}
\newcommand{\vcross}{{\boldsymbol{\times}}}
\newcommand{\grad}{\mbox{\boldmath$\nabla$}}
\newcommand{\thth}{\hspace{1.5pt}}
\newcommand{\curl}{\grad\vcross}
\newcommand{\Curl}{\grad\vcross\thth}
\renewcommand{\div}{\grad\vdot}
\newcommand{\Div}{\grad\vdot\thth}
\newcommand{\boldv}{{\mathbf{v}}}
\newcommand{\boldk}{{\mathbf{k}}}
\newcommand{\kperp}{k_{\scriptscriptstyle\!\perp}}
\newcommand{\Alfven}{Alfv{\'e}n}



%
%

%
%


\begin{abstract}
The importance of the chromosphere in the mass and energy transport within the solar atmosphere is now widely recognised. This review discusses the physics of magnetohydrodynamic (MHD) waves and instabilities in large-scale chromospheric structures as well as in magnetic flux tubes. We highlight a number of key observational aspects that have helped our understanding of the role of the solar chromosphere in various dynamic processes and wave phenomena, and the heating scenario of the solar chromosphere is also discussed. The review focuses on the physics of waves and invokes the basics of plasma instabilities in the context of this important layer of the solar atmosphere. Potential implications, future trends and outstanding questions are also delineated.
\end{abstract}


%
%

%


%
%
%
%

%
%
\section{Introduction}
The complex magnetic field of the solar chromosphere exerts significant effects on the propagation and dissipation of magnetohydrodynamic (MHD) waves. The chromosphere channels mechanical energy from the photosphere into the transition region (TR) and corona, leading to several physical processes such as wave reflection and 
mode-conversion, which depend on the magnetic field strength and the local plasma properties (e.g., Vecchio et al., 2007; Srivastava et al. 2008a; Cally and Khomenko 2019; Ballester et al. 2020). In the quiet-Sun the magnetic field is rooted in the intergranular lanes forming intense flux tubes  which are subjected to continuous motions of their footpoints. 
The physical nature of the drivers near the footpoints of the flux tubes results in a variety of wave modes which evolve as they travel through the solar photosphere and lower chromosphere. Such MHD waves may also be evolved {\it in situ} in the chromosphere and can propagate into TR and inner corona imparting some Poynting energy flux to overcome their radiative losses, as recently observed in a variety of chromospheric structures (e.g., Kukhianidze et al. 2006; Zaqarashvili et al. 2007; Jess et al. 2009; Morton et al. 2012; Kuridze et al. 2013; Srivastava et al. 2017; Jess et al. 2020).
For example these MHD modes include: (i) kink waves excited by the horizontal buffeting motions, (ii) slow waves due to pressure fluctuations, (iii) torsional Alfv\'en waves generated by the twisting motions; or/otherwise important in the long-wavelength limit 
fast and slow magnetoacoustic-gravity waves as well as their 
hybrid consisting of coupled Alfv\'en and magnetoacoustic-gravity waves (e.g., Ulmschneider et al. 1991; Hasan et al. 2003; De Pontieu et al. 2007; Khomenko et al. 2008a; Jess et al. 2009; McIntosh et al. 2011; Morton et al. 2012; Mathioudakis et al. 2013; Jess et al. 2015; Srivastava et al. 2017; Liu et al. 2019).

The MHD waves and oscillations are also associated with the active region sunspots. The chromospheric plasma, which is a low-lying and rather strongly magnetized region above these spots, possesses exotic physical processes and their subsequent effects on the wave propagation. 
Wave excitation mechanisms above these spots, mode conversion, the formation of shocks, frequency dependent reflection from the transition region, the effects of complex magnetic structuring and inhomogeneities in the penumbra, put forward a multitude of physical processes that affect the evolution and dissipation of MHD wave modes and determine their role in chromospheric and coronal heating (e.g., Crouch and Cally 2005; Botha et al. 2011; Khomenko and Cally 2012; Tian et al. 2014; Khomenko and Collados 2015; Grant et al. 2018; Srivastava et al. 2018; Kang et al. 2019; Cho and Chae 2020). While the complex structuring of plasma and magnetic field above sunspots influence highly the evolutionary and dissipative properties of MHD waves over the large-scale environment at long-wavelength limits, there is an ample  evidence that these strongly magnetized 
structures can act as a MHD wave-guide 
for the excited tubular modes (e.g., kink, sausage, and torsional modes) which transport
the significant amount of energy upwards into the overlying corona possibly for its heating (e.g., Grant et al. 2015; Moreels et al. 2015a; Jess et al. 2017; Keys et al. 2018). 

The plasma of solar prominences have properties similar to the properties of the chromospheric plasma, apart from the superstrong magnetic field in the intense flux tubes.
These structures are treated as thermally and pressure isolated, enveloped inside a Prominence-Corona-Transition-Region (PCTR) and are made-up of partially ionized plasma (Parenti 2014). It has been proposed that the excitation and dissipation of MHD waves in prominences can contribute to their heating in addition to the more dominant radiative heating (Arregui 2015). Complex and turbulent flows, strong inhomogeneities in plasma and magnetic field properties under the presence of gravity also yield some typical instabilities in such isolated chromospheric plasma systems, e.g., Rayleigh-Taylor (R-T), Kelvin-Helmholtz (K-H) instabilities (e.g., Berger et al. 2008; Ryutova et al. 2010; Berger et al. 2010; Hillier et al. 2011; Innes et al. 2012; Berger et al. 2017; Mishra et al. 2018; Mishra and Srivastava 2019). As a result of the existence of waves and a wide variety of plasma flows induced by their nonlinear behaviour
in the prominences, which are strongly influenced by
the large magnetic Reynolds numbers of the ambient system, such solar prominences may develop turbulence that further contribute significantly to the heating rate of these large-scale structures (Hillier and Polito 2018).

The solar chromosphere is different from the corona as far as MHD wave modes, instabilities, and heating are concerned. It requires higher amounts of energy input ($10^{6}$--$10^{7}$ erg cm$^{-2}$ s$^{-1}$) to balance its radiative losses compared to the solar corona ($10^{4}$--$10^{6}$ erg cm$^{-2}$ s$^{-1}$). 
Realistic models of the chromospheric plasma involve a multi-fluid approach with a finite plasma-beta and accommodate the additional physical effects  of partial ionisation and radiative transfer in non-local thermodynamic equilibrium (e.g., Hansteen et al. 2007; Soler et al. 2012; Mart{\'\i}nez-Sykora et al. 2015; Soler et al. 2019; Ballai et al. 2019). Although the single continuum fluid consideration at MHD length and time scales will suffice in dealing with the various wave modes or instabilities in the solar chromosphere, we do not invoke some additional frequency dependent physics of the region, e.g., wave damping by ion-neutral collisions, etc., there  (Ballester et al. 2018).

Alternatively, the direct dissipation of electric current facilitated by the  magnetic reconnection provides a stringent physical scenario about another significant mechanism to heat the solar atmosphere (e.g., nano-flare heating; bulk plasma heating; liberating energy through twists and braiding of the magnetic field lines, etc.) (e.g., Cargill and Klimchuk 2004; Winebarger et al. 2013; Klimchuk 2015; Xue et al. 2016; Srivastava et al. 2019). Apart from the magnetic reconnection generated heating, when we consider the two- or multi-fluid scales and relevant physical scenario especially in the context of the solar chromosphere, regardless of the waves the frictional heating can be significant as it simply accounts for the case where the ions are moving relatively to the neutrals and the friction between them can be significant (e.g., Al Shidi et al. 2019). The similar physical scenario is true for the waves and instabilities also in the frame-work of the solar chromosphere where the deviation from MHD scales to fluid scales lead the additional physics (e.g., ion-neutral collisions; ambipolar diffusion, etc) leading the evolution and dissipation of the perturbations at much smaller spatial scales (e.g., Soler et al. 2012; D{\'\i}az et al. 2012; Soler et al. 2013; Zaqarashvili et al. 2013; Khomenko et al. 2014a; Soler et al. 2015a,b; Khomenko and Collados 2012; Mart{\'\i}nez-G{\'o}mez et al.2018). Complementing the alternative physical scenario significant for the chromospheric heating and multitude of its dynamical plasma processes, we conjecture herewith that the present article review the role of the waves and instability in MHD environment of the solar chromosphere.

There have been several reviews that discuss the key observations and theoretical developments on MHD waves in the solar chromosphere and corona and their contribution to heating (e.g., Narain and Ulmschneider 1996; Zaqarashvili and Erd{\'e}lyi 2009; Mathioudakis et al. 2013; Jess et al., 2015; Arregui 2015; Aschwanden 2019; Van Doorsselaere et al. 2020).  
In the present review, we aim to discuss the physics of MHD waves and instabilities in the 
chromospheric plasma, and their heating capabilities. The description of shocks is also included in the 
context 
of chromospheric heating. Key updates on new observational results of these physical phenomena are also elucidated. In Section 2, we address the physical behaviour of MHD waves in the large-scale chromosphere, and depict recent progress on this important topic. Section 3 describes the physics and current trend of the research of various MHD waves in the structured chromosphere and its magnetic fluxtubes. We demonstrate the overall development in the understanding of solar chromospheric heating in Section 4. Thermally isolated chromospheric plasma structures, i.e., solar prominences, are illustrated in Section 5. We discuss the evolution and observational aspects of instabilities that may have significant impact on the chromospheric plasma systems, including their contribution to heating and dynamics. In the last section, we present discussion and conclusions. Here, we also list a number of outstanding questions that could be addressed with the use of existing high-resolution and upcoming ultra high-resolution ground (e.g., 4m-DKIST, upcoming 4m-EST, 2m-NLST) and space (e.g., SDO, IRIS, Parker Solar Probe, Solar Orbiter, upcoming Aditya-L1, Solar-C, etc) borne observations, which will also ensure one-to-one continuous refinements of the theoretical understanding.

%
%
\section{Magnetohydrodynamic Waves in the Large-Scale Chromosphere}\label{Sec:largescale}
%
MHD waves take their simplest linear form in a uniform ideal fully ionized magneto\-fluid, providing three restoring forces: gas pressure, magnetic pressure, and magnetic tension. The Alfv\'en wave, in its pure form, is driven by tension alone (though see the discussion in Section~\ref{sec:torsional_waves} for a more nuanced view), whilst the two magneto\-acoustic waves, fast and slow, result from the combination of all three.

However, the solar chromosphere is very far from simple. To these basic forces, we must add the effects of magnetic structure, including current sheets, 
gravitational stratification, giving rise to refraction, reflection, steepening and buoyancy; nonlinearity; partial ionization; radiation; non-LTE; and substantial background flows. Under these conditions, the simple fast/slow/Alfv\'en categorization may be inadequate, or at least not globally valid.

There are also various sources of the wave motions, such as the Sun's internal oscillations, the p-modes; granulation at the photosphere; small reconnection events, etc.\footnote{The net horizontal velocities in granules at the photosphere are of order 0.5 $\rm km\ s^{-1}$, with the corresponding vertical velocities being about half that (Mattig et al. 1981). For comparison, the combined p-mode oscillation velocity amplitude in the five-minute range is also about 0.5 $\rm km\ s^{-1}$, mostly vertical, though this is made up of millions of individual modes whose individual amplitudes do not exceed 10~$\rm cm\ s^{-1}$ (Christensen-Dalsgaard 2002).} A commonly invoked wave driver is photospheric granulation. Although it has been argued that granulation is a part of a Kolmogorov turbulent spectrum (Espagnet et al. 1993),  this has been strenuously disputed by Nordlund et al. (1997) on the basis that the turbulence is non-steady and full properties cannot be inferred from a single snapshot. Nevertheless, a granular scale of 1--2 Mm is prominent, and timescales of a few minutes are typical.
Cranmer and van Ballegooijen (2005) famously based a whole-heliosphere Alfv\'en wave model on such granular and supergranular driving with peak `kink-mode wave energy' power at periods of 5--10 minutes, inferred from observed Magnetic Bright Point (MBP) motions.

The Sun's internal normal modes, the p- and f-modes, have a similar timescale, peaked around five minutes, though their horizontal length scales are far larger. A typical granule size of 1 Mm corresponds to a very high spherical harmonic degree $\ell\approx4400$, but nearly all p-mode power lies at much smaller $\ell$, so these drivers of atmospheric waves are quite distinct.

Using the Coronal Multi-Channel Polarimeter (CoMP) to observe a region of bright active region coronal loops in the FeXIII 1074.7 nm emission line, Tomczyk et al. (2007) detected a distinct peak in oscillatory power around 3.5 mHz, which they associated with the Sun's internal p-modes. The power peak correlates well with both the position and width of the average power spectrum of intermediate degree photospheric seismic oscillations. These coronal oscillations were initially identified as Alfv\'en waves, but are more likely kink magneto\-acoustic waves (Erd{\'e}lyi and Fedun 2007; Van Doorsselaere et al. 2008) if sufficient flux tube structuring is available to support them.

It is also possible that the 3.5 mHz peak is characteristic of coronal loop resonant frequencies rather than p-modes. Nechaeva et al. (2019) identify a large number of decaying kink oscillations with periods in the range of roughly 2 to 26 minutes (their Table 1), i.e., frequencies between 0.6 and 8 mHz. See also the proposed self-oscillatory mechanism of Nakariakov et al. (2016). However, this frequency range seems much broader than the CoMP observations suggest. Their persistence across all coronal magnetic field topologies, even in open field, argues in favour of p-mode origins; see Morton et al. (2015), Morton et al. (2016) and Morton et al. (2019) for further details. Despite these indications, the role of p-modes in driving coronal oscillations remains an open question.

However, if p-modes are the cause, this poses the question of how these Alfv\'en or Alfv\'enic (near transverse and near incompressive) waves propagate through the chromosphere, and how they penetrate the transition region with enough amplitude to explain the observed transverse coronal oscillations (McIntosh et al. 2011; McIntosh and De Pontieu 2012).

The energy carried by these coronal waves was estimated by McIntosh et al. (2011) and McIntosh and De Pontieu (2012), using more resolved observations from \emph{Hinode}/SOT and SDO/AIA, to
provide substantial coronal heating and solar wind acceleration. However, this was based on wave flux formulae relevant to Alfv\'en waves in a uniform plasma. It is now more widely believed that the observed oscillations are kink waves on discrete flux tubes, for which filling-factor considerations reduce the calculated fluxes by factors of 10--50 (Goossens et al. 2013a; Goossens et al. 2013b).

Waves in and on magnetic structures such as spicules may contribute 
to coronal heating and solar wind origin, but in this section the focus is on the large scale in which the horizontal length scales are far larger than the typically 100--200 km vertically imposed by gravity. Due to the very weak ionization levels in the photosphere (typically of order $10^{-4}$) there has been some doubt that Alfv\'en waves could be directly excited there to any significant amplitude (Vranjes et al. 2008), though this has been disputed. The knub of the discrepancy is that only the ionized fluid is given any initial velocity in the analysis of Vranjes et al. (2008), whereas Tsap et al. (2011) effectively give the ionized and neutral fluids the same initial velocity. Energy is therefore spread very thinly when it is used to accelerate the predominant neutral fluid in the model of Vranjes et al. (2008), thereby greatly reducing amplitude (Soler et al. 2013).

In any case, the p-mode component of any atmospheric wave driving in quiet Sun is predominantly acoustic and vertical at the photosphere, and clearly not Alfv\'enic. How does this ultimately register as transverse waves in the transition region and corona? Let us discuss the route taken by initially acoustic waves in traversing the chromosphere.

Two characteristic frequencies associated with an atmosphere determine the behaviour of acoustic and acoustic-gravity waves: the Brunt-V\"ais\"al\"a or buoyancy frequency $N$ defined by $N^2=g/H-g^2/c_s^2$, where $g$ is the gravitational acceleration, $H$ is the density scale height, and $c_s$ is the sound speed, and the acoustic cutoff frequency $\omega_c$. 

Acoustic cutoff frequency is a very fragile quantity, with different formulae resulting from using different variables in describing the equations (Schmitz and Fleck 1998; Schmitz and Fleck 2003). In a non-magnetic plane-stratified atmosphere, introducing dependent variable $\Psi=\rho^{1/2}c_s^2 \Div\boldv$ (Deubner and Gough 1984) and assuming $\exp[i(k_xx+k_yy-\omega t)]$ dependence, the linearized wave equations may be reduced to the form
\begin{equation}
\frac{d^2\Psi}{dz^2}+
\left(
\frac{\omega^2 -\omega_c^2}{c_s^2} +\frac{N^2}{\omega^2}k_h^2-k_h^2
\right)\! \Psi= 0,   \label{AGeqn}
\end{equation}
where $k_h=\sqrt{k_x^2+k_y^2}$ is the horizontal wave number, 
and 
$\omega_c^2 = (c_s^2/4H^2)\left(1-2H'\right)$. Although mathematically elegant, the form of $\omega_c$ in this formulation is a computational nightmare when used for calculations in a tabulated empirical atmosphere such as the widely used VAL C (Vernazza et al. 1981) (a tabulation of proton density, electron density, pressure, temperature, etc., in a one-dimensional stationary quiet chromosphere model based on non-LTE radiative transfer and many atomic and ionic species). It yields very noisy results because of the second derivative of density implicit in the $H'$ term. Nevertheless, in an isothermal atmosphere, where $H'=0$, the unambiguously correct `isothermal acoustic cutoff' $\omega_{ci}=c_s/2H$ is recovered.

Identifying the term in the bracket in Eq.~(\ref{AGeqn}) as the square of the vertical wavenumber $k_z$, the dispersion relation for these non-magnetic waves is $c_s^2k_z^2=\omega^2(\omega^2-\omega_c^2)-(\omega^2-N^2)c_s^2k_h^2$,
which allows propagation vertically only if the right hand side is positive, i.e., if
\begin{equation}
\omega^2 
\gtrless 
  \frac{1}{2} \left[
    c_s^2k_h^2+\omega_c^2\pm\sqrt{(c_s^2k_h^2+\omega_c^2)^2-4c_s^2N^2k_h^2}
  \right],
\end{equation}
corresponding respectively to acoustic waves (upper signs $>$ and $+$) and gravity waves (lower signs $<$ and $-$). Between the two, the waves are vertically evanescent and do not propagate. In particular, if $k_h=0$, propagation occurs only for frequencies $\omega>\omega_c$, which explains the name `cutoff frequency'. Typically, the acoustic cutoff frequency in the solar chromosphere is around 5 mHz, so by rights, sound waves should not be able to propagate upward below this frequency. 

This provides an opportunity for trapping waves between two heights, corresponding to a cavity. A 3-minute chromospheric cavity is reasonably well-established in sunspot atmospheres (e.g., Botha et al. 2011; Jess et al. 2020), although different interpretations can often be provided by considering line formation heights (Bogdan and Judge 2006). There is evidence of a very small number of trapped `chromospheric eigen modes' in the quiet Sun too (Deubner et al. 1996),
though Fleck and Schmitz (1991) attribute 3-min chromospheric oscillations to resonant excitation of an acoustic cutoff frequency mode that exists even in isothermal models where the acoustic cutoff frequency is constant. On the other hand, the chromosphere may be so dynamic that discussion of cavities is moot (Carlsson and Stein 1998). 

However, the chromosphere is magnetic (Carlsson and Stein 1995). The acoustic cutoff effect is ameliorated by sufficiently strong magnetic field (Bel and Leroy 1977), allowing propagation with a travelling (non-evanescent) vertical component if $\omega > \omega_c \cos\theta$, where $\theta$ is the inclination angle of the magnetic field from the vertical. This opens `magnetic portals' (Jefferies et al. 2006) in the low atmosphere that allow waves well below 5 mHz to propagate into the upper atmosphere. This is sometimes called the `ramp effect', and is most effective in quiet Sun in strong small scale magnetic elements such as supergranular network where $\beta\lesssim 1$. Of course, it is very effective in active regions.

A crucial insight is that the three MHD wave types: slow, Alfv\'en and fast, are not fixed across inhomogeneous atmospheres. They may be unambiguously one of these types over much of the atmosphere, but change to another type where the phase velocities of the two waves nearly match. This became apparent with the exact 2D solutions for ideal magneto\-acoustic waves in an isothermal gravitationally stratified atmosphere with uniform inclined magnetic field (Zhugzhda and Dzhalilov (1984) in terms of Meijer-G functions, or Cally (2009) in terms of the more elementary ${}_2F_3$ hypergeometric functions), where exact fast/slow coupling coefficients may be found. The Alfv\'en wave is strictly decoupled in 2D where the wave propagates in the plane defined by gravity and the magnetic field.

Two rather different types of mode conversion play a significant role. The first is fast/slow conversion at any Alfv\'en-acoustic equipartition surface $v_A=c_s$, where $v_A$ is the Alfv\'en speed. In the quiet solar chromosphere this might typically be at heights of a few hundred kilometres, though in sunspots it can be below the photosphere. Above this region the Alfv\'en speed increases roughly exponentially, due to the exponential decrease in density with height, whilst the sound speed varies only slowly until the transition region is reached.

\begin{figure}[p]
\begin{center}
\includegraphics[width=\textwidth]{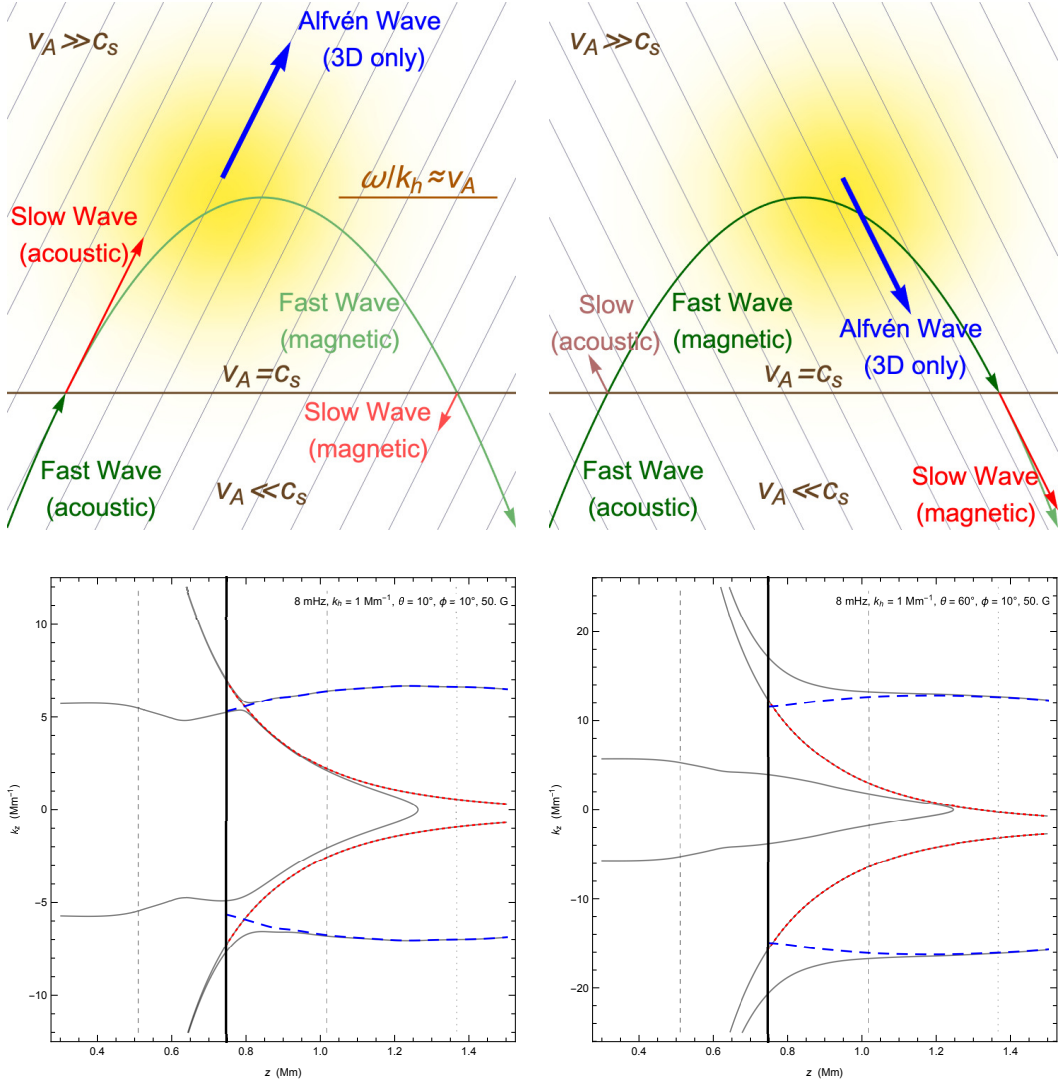}
\caption{Top left: schematic diagram of an injected fast (acoustic) ray incident from bottom left travelling upward and passing through the Alfv\'en-acoustic equipartition level $v_A=c_s$, mostly transmitting as a slow wave but partly converting to a fast wave. The fast wave reflects near where its horizontal phase speed matches the Alfv\'en speed, where it partially converts to an upgoing Alfv\'en wave (3D only). Top right: same, except the magnetic field is inclined contrary to the propagation direction, so most energy at $v_A=c_s$ goes into the fast wave, with subsequent conversion to a downward Alfv\'en wave. Bottom left: $z-k_z$ dispersion diagram for an 8 mHz, $k_h=1$ $\rm Mm^{-1}$ wave in uniform 50 G magnetic field inclined $\theta=10^\circ$ from the vertical and oriented $\phi=10^\circ$ out of the $x-z$ plane. The ray enters at the bottom on the fast branch with $k_z\approx 6$ $\rm Mm^{-1}$, reaches $v_A=c_s$ around $z=0.75$ Mm (thick vertical line), and passes very close to the slow branch. Partial mode conversion happens at this avoided crossing. The subsequent fast ray continues upward and reflects around $z=1.3$ Mm, but experiences a long near-correspondence with the intermediate (Alfv\'en) mode and again partially converts. The blue dashed line indicates the asymptotic $v_A\gg c_s$ slow (acoustic) wave and the red dashed curve is the asymptotic Alfv\'en wave. The atmosphere is VAL C. Bottom right: the same, but with $\theta=60^\circ$, for which case the fast-Alfv\'en conversion occurs much closer to the fast wave apex. The upper part of this figure is adapted from Khomenko and  Cally (2012).}
\label{fig:PSC}
\end{center}
\end{figure}

Fast/slow conversion is a very compact process. In a 2D scenario, Schunker and Cally (2006) show that the energy transmission coefficient is (to a good approximation) given by 
\begin{equation}
T = \exp\left[
-\frac{\pi h k^2 \kperp^2}{|k_z| (k^2+\kperp^2)}
\right]_{v_A=c_s}
=  \exp\left[
-\frac{\pi h k^2 \sin^2\alpha}{|k_z| (1+\sin^2\alpha)}
\right]_{v_A=c_s}
,    \label{T}
\end{equation}
where $k=|\boldk|$ is the total wave number, $\kperp$ is the component of wave vector $\boldk$ perpendicular to the magnetic field, $h=[d(v_A^2/c_s^2)/dz]^{-1}$ is the equipartition layer scale height, and $\alpha$ is the attack angle between the wave vector and the magnetic field (so $\kperp=k\sin\alpha$). The corresponding conversion coefficient is $C=1-T$ (ignoring the phase). Note that transmission is total ($T=1$) for zero attack angle, meaning that a field-aligned fast wave from $v_A<c_s$ (making it predominantly acoustic in nature) passes perfectly through to $v_A>c_s$ still as acoustic wave (now the slow wave), with no conversion to a magnetically dominated (fast) wave. On the other hand, at non-zero attack angle, the incident wave splits into a combination of slow and fast waves. This is seen in numerous simulations (e.g., Felipe et al. 2010), and even persists for shock waves (Pennicott and Cally 2019). Figure \ref{fig:PSC} illustrates the two mode conversion processes both schematically (upper) and through $z-k_z$ dispersion diagrams for representative cases. 

The other important mode conversion process is fast-to-Alfv\'en, which only occurs in 3D where gravity, the magnetic field lines and the wave vector are not co-planar. This discussion is predicated on the assumption that vertical gravitational stratification, typically with scale height of order 150 km in the chromosphere, is the dominant inhomogeneity. It would be altered in the presence of similar or smaller horizontal magnetic scales, though the general principles persist. High in the chromosphere, after (magnetically dominated) fast waves have been generated by mode conversion at $v_A=c_s$ they propagate rapidly up into the region where $v_A\gg c_s$ and their dispersion relation is $\omega^2\approx v_A^2(k_h^2+k_z^2)$. They then reflect ($k_z=0$) at around where the horizontal phase speed equals the Alfv\'en speed. It is near this point, but typically distributed over several scale heights, that the fast wave partially loses energy to generate an Alfv\'en wave, that either travels upwards along the field lines if those field lines are inclined in the direction of propagation, or downward if the field is inclined counter to the direction of wave propagation (Cally and Hansen 2011). Both mode conversion processes are confirmed computationally by Cally and Goossens (2008). Fast-to-Alfv\'en conversion is found to enhance Alfv\'en wave penetration of the transition region compared to Alfv\'en waves generated at the photosphere (Hansen and Cally 2012).

These two mode conversion processes are confirmed by simulations in model sunspots (Khomenko and Cally 2012; Felipe 2012). By its very nature, mode conversion is not something that can be observed locally with telescopes and instruments, since the process happens where elsewhere-distinct wave types (fast, slow, Alfv\'en) become almost degenerate and indistinguishable. In the language of Goossens et al.(2019), they exhibit `mixed properties'. This makes simulation particularly valuable in verifying and exploring the process in realistic atmospheres. From an observational viewpoint, mode conversion is best observed through its effects: the disappearance of p-modes hitting sunspots (Braun et al. 1987); the signatures of transverse waves in the corona associated with the 5-minute oscillations (Tomczyk et al. 2007); the acoustic halos surrounding sunspots that bear the hallmarks of returning fast waves from above (Khomenko and Collados 2009; Rajaguru et al. 2013; Rijs et al. 2016); the influence of magnetic field inclination on chromospheric waves (Rajaguru et al. 2019); etc. As far as the chromosphere goes, the \emph{effects} of mode conversion are potentially profound, providing rungs of a ladder for energy propagation from the photosphere to the corona. More detail about links between mode conversion and chromospheric wave heating may be found in Section \ref{Sec:heating}. 

In summary, above the acoustic cutoff frequency (about 5 mHz), acoustic waves from the surface or sub-surface, including those due to p-modes, may partially convert to magnetically dominated fast waves at $v_A=c_s$, and then, although these fast waves reflect and do not reach the corona, they may pass on considerable energy to Alfv\'en waves, and these can penetrate the transition region. However, below $\omega_c$ they would need to enter via magnetic portals in inclined intense supergranular network flux elements, where they will encounter an $v_A=c_s$ equipartition layer much lower in the atmosphere than in the broader canopy (Bogdan et al. 2003), perhaps several times as the Alfv\'en speed may decrease with height due to rapid field-line spread before increasing again as the density decreases. The picture is complicated further by a significant fraction of internetwork field (around half) that also joins the network field that reaches the corona (Schrijver and Title 2003). This three-stage process may help explain the observational link between the Sun's sub-surface seismic wave field and the ubiquitous Alfv\'enic oscillations in the corona (Morton et al. 2015).

We have seen that the Sun's internal normal modes, the p- and f-modes, can drive chromospheric oscillations, and perhaps even be detected in the corona. It is also well-established that the chromosphere affects the subsurface global seismology, particularly as activity varies across a solar cycle. Models are explored by Campbell and Roberts (1989), Evans and Roberts (1991), and reviewed at length by Erd{\'e}lyi (2006), who asks `Can the tail wag the dog?'. Furthermore, chromospheric waves can affect internal local seismology via a combination of the mechanisms discussed above, especially in and around active regions. Time-distance analysis of simulations by Cally and Moradi(2013) reveals that significant `travel time' discrepancies of up to 40 sec are introduced around active regions by reflected and mode-converted waves re-entering the subphotosphere, thereby polluting the wave field and making it more difficult to infer subsurface features from local seismic data. Rijs et al. (2016) also demonstrate via simulations that the observed acoustic halo around active regions is due to fast waves returning from the chromosphere. 

A further feature of the photosphere and low chromosphere that affects wave propagation is the very low ionization fraction there, as low as $10^{-4}$. This has several effects. First, by increasing the mean molecular weight it profoundly alters the hydrostatic pressure and density structure of the equilibrium atmosphere, compared to a fully ionized model. Second, it introduces further physics, notably the Hall effect and ambipolar diffusion (Khomenko et al. 2014b). Although the Hall term is conventionally written in terms of a diffusivity $\eta_H=B/(e n_e \mu)$, where $e$ is the elemental charge, $n_e$ is the electron number density and $\mu$ is the magnetic permeability, it is not in fact diffusive (it does not contribute to the thermal energy). Instead, it induces a continuous oscillation between fast and Alfv\'en waves in the form of a precession of the polarization (Cally and Khomenko 2015). However, the dimensionless Hall parameter for a wave of frequency $\omega$, $\epsilon=\omega/(f \Omega_i)$ where $f=\rho_i/\rho$ is the ionization fraction and $\omega_i$ is the ion gyroradius, is insignificant in the solar chromosphere for all but weak fields (a few Gauss) and very high frequencies (of order 1 Hz or above). Simulations by Gonz{\'a}lez-Morales et al. (2019) confirm that Alfv\'en waves are indeed generated by Hall coupling at these high frequencies, and reach amplitudes sufficient to play a role in coronal heating.

In one-fluid descriptions, the ambipolar diffusivity is $\eta_A=\xi_n^2 B^2/(\alpha_n\mu)$, where $\xi_n=\rho_n/\rho=1-f$ is the fractional contribution of the neutrals to the density and $\alpha_n$ is the neutral collisional parameter (a density-weighted collision frequency of neutrals with electrons and ions). Linear and nonlinear simulations of 10 mHz waves in an inclined rectangular packed ensemble of flux tubes with an internal/external field strength contrast of a few percent (Khomenko and Cally 2019) reveal that linear scattering greatly enhances the production of Alfv\'en waves compared to the horizontally uniform case; that nonlinear effects become important in the upper chromosphere; that the flux tube structuring enhances Poynting flux reaching the corona by about 35\% and reduces reflection at the transition region by 50\%; and that ambipolar diffusion, though notionally weak at this frequency, does have some effect in thermalizing a small fraction of fast waves in the chromosphere (see also Cally and Khomenko 2019) due to the steep gradients that develop there in the magnetically structured model, though this is not fully resolved numerically (10 km) and may in reality be stronger.  Recently, Al Shidi et al. (2019) have shown that ion-neutral frictional heating can be significant in the chromosphere, and lead to the generation of jets.

The influence of partial ionization on chromospheric waves is not restricted to processes in the chromosphere only. Gonz{\'a}lez-Morales et al. (2020) find that ambipolar diffusion near-surface battery-excited dynamo reduces the flux of Alfv\'en waves generated, whilst the Hall effect substantially enhances chromospheric Alfv\'en fluxes. Heating aspects of waves in the large-scale solar chromosphere is discussed in details in Section~4.

In the next-section, we review the evolution of the waves in the fine structured chromospheric flux tubes.
\section{Waves in Structured Flux tubes in the Chromosphere}\label{Sec:structured}

In Section~\ref{Sec:largescale}, we have described the detailed physics of the slow, fast, and Alfv\'en waves in the homogeneous and unbounded solar plasma, and discussed especially the
role of the solar chromosphere on their propagation. In the present Section~\ref{Sec:structured}, we discuss the physics of MHD wave modes (e.g., kink, sausage, and torsional waves) in the fine structured tubes, and the recent trends of research in the frame-work of the solar chromosphere. 

\subsection{Various MHD modes and kink waves in structured non-uniform flux tubes}\label{Sec:kink}

Kink waves in cylindrical plasmas are non-axisymmetric waves that correspond in a system of cylindrical coordinates to the azimuthal wave number $m=1$. They are important because they displace the flux tube as a whole. This phenomenon is  well-known for uniform flux tubes with piecewise constant density. When the flux tube is non-uniform there are strong rotational motions in addition to this translational motion. The starting point of its physical description can be two papers from 1970s by Ionson (1978) and Wentzel (1979) and a 
book by Hasegawa and Uberoi(1982). Wentzel (1979) studied  hydromagnetic surface waves on cylindrical flux tubes. He pointed out that these surface waves differ qualitatively and quantitatively from ordinary magnetohydromagnetic waves (e.g., Alfv\'{e}n, slow, fast waves; Sect.~\ref{Sec:largescale}). Wentzel solved the linear MHD equations for arbitrary azimuthal wave number $m$. The $m=1$ corresponds to kink waves. Wentzel (1979) recalled  that Ionson (1978) proposed a theory for heating coronal loops by Alfv\'{e}nic surface waves. He derived the waves for a plane surface and assumed $k_y = 1/R$ for a cylinder. The $k_y = 1/R$ corresponds to $m=1$ in cylindrical geometry, i.e. kink waves. 
Resonant absorption requires smooth non-uniformity. It does not work for discontinuous profiles.
Hasegawa and Uberoi(1982) emphasized that $P'$ (pressure perturbation) plays a special role for MHD waves in a non-uniform plasma. They noted that "The basic characteristic of the ideal Alfv\'{e}n wave is that the total pressure in the fluid remains constant during the passage of the wave as a consequence of the incompressibility condition. For  inhomogeneous medium, however, the total pressure, in general, couples with the dynamics of the motion, and the assumption of neglect of pressure perturbations becomes invalid."  Often the results for MHD waves on a uniform plasma of infinite extent are used as guidelines. The MHD waves can be separated in Alfv\'{e}n waves and magnetoacoustic waves. The Alfv\'{e}n waves propagate parallel vorticity and are incompressible. The magnetoacoustic waves are compressible and do not propagate parallel vorticity.  However when the equilibrium configuration is changed either by a discontinuous variation of the equilibrium quantities or by a continuous variation in a non-uniform layer, deviations from this classification occur.
%
%
%

After this very fundamental illustration, now we focus on the linear MHD waves superimposed on a 1-D cylindrical plasma column of radius $R$ in static equilibrium. It basically describes a plasma cylinder of a circular cores-section, extended along the field, and smoothly non-uniform in the radial direction. The equilibrium density $\rho_0(r)$, equilibrium pressure $p_0(r)$ and the components of the equilibrium magnetic field $B_{z,0}(r), B_{\varphi,0}(r)$ are functions of $r$ or constant. Since the equilibrium quantities are independent of $\varphi$ and $z$ the wave variables can be put proportional to the exponential factor $\exp (i (m\varphi + k_{z}z))$ with  $m,k_z$ the azimuthal and axial wave numbers, $m$ is an integer. 
Kink waves correspond to $m=1$. 

\noindent The  linear MHD waves superimposed on this 1-D cylindrical plasma column  can be described by two  ordinary differential equations for the radial component of the Lagrangian displacement $\xi_r$ and the Eulerian perturbation of total pressure $P'$ (Sakurai et al. 1991). The components of the Lagrangian displacement in the magnetic surfaces perpendicular and parallel to the magnetic field lines  $\xi_{\perp},\;\xi_{\parallel}$,  compression  $ \nabla \cdot \vec{\xi} $  and vorticity  $(\nabla \times \vec{\xi}) $ can be  given by expressions in terms of $\xi_r$ and $P'$ and their derivatives. Algebraic expressions for $\xi_{\perp},\;\xi_{\parallel}, \nabla \cdot \vec{\xi} $ can be found in Sakurai et al. (1991).  Equations for the components of $(\nabla \times \vec{\xi})$ can be found in Goossens et al. (2019). The components of vorticity are in general non-zero.  All of the wave variables are coupled. The MHD waves have mixed properties, they propagate both compression and parallel vorticity and have non-zero radial, perpendicular and parallel  components of displacement and vorticity. A situation in which a subset of the wave variables is not coupled to the other wave variables is an exception. Such a situation appears for axisymmetric motions in the presence of a straight field. In general, the clear division into Alfv\'{e}n waves and magneto-sonic waves  that exists for a uniform plasma of infinite extent does not any  longer hold. 

\noindent  Let us now focus on MHD waves in presence of a straight field. For a straight field  ($B_{\varphi,0} = 0\;$) the magnetic surfaces are cylinders: $r = \mbox{constant}$. The  $\varphi -$ and   $z -$ directions are the directions in the magnetic surfaces respectively perpendicular and parallel to the magnetic field lines. The $r -$direction is normal to the magnetic surfaces.  
The equations to be used are  e.g. equations (45) of Goossens et al. (2019)  for the components of the displacement $\vec{\xi}$ and compression $\nabla \cdot \vec{\xi}$ 

\begin{eqnarray}
\xi_r & = & \frac{1}{\rho_0 (\omega^2 - \omega_A^2)} \frac{d P'}{dr}, \nonumber  \\
&& \nonumber \\ 
\xi_{\perp} & = & \xi_{\varphi} = i \frac{m}{r}\;
\frac{1} {\rho_0 (\omega^2 - \omega_A^2)} \;P', \nonumber  \\
&& \nonumber  \\
\xi_{\parallel} & = & \xi_z = i k_z \;\frac{c_s^2}{c_s^2 + v_A^2}\; \frac{1}{ \rho_0 (\omega^2 - \omega_C^2) } P' , \nonumber \\
&& \nonumber  \\
\nabla \cdot \vec{\xi} & = & \frac{-\omega^2 \;  P' }  { \rho_0 (c_s^2 + v_A^2)(\omega^2 - \omega_C^2)}
\label{EqStrFieldA}
\end{eqnarray}

The local Alfv\'{e}n frequency $\omega_A\;$ and the  local cusp frequency $ \;\omega_C$  are  defined as 
\begin{equation}
 \omega_A^2  = \;k_z^2\;v_A^2 = k_{\parallel} \;v_A^2, \;\;
 \omega_C^2  = \frac{c_s^2}{c_s^2 + v_A^2}  \omega_A^2   
\label{Alfven-CuspFre1}
\end{equation}

In a non-uniform plasma $\omega_A$ and $\omega_C$ are function of position. For a given set of wave numbers $(m, k_z)$ they   map out two  ranges of frequencies known as the Alfv\'{e}n continuum and the cusp continuum. The $\; v_A$, $c_s$ are the velocity of sound and the Alfv\'{e}n velocity.
\begin{equation}
 v_A^2 = \frac{\displaystyle B_0^2}{\displaystyle \mu \; \rho_0}, \;\;
c_s^2 = \frac{\displaystyle \gamma p_0}{\rho_0}
\label{VAVS}
\end{equation}

\noindent For axi-symmetric motions with $m=0$ the equation for $ \xi_{\perp} =\xi_{\varphi}$ 
\begin{equation}
(\omega^2 - \omega_A^2)\; \xi_{\varphi} =0
\label{AxiAW}
\end{equation}
is  decoupled from the remaining equations. The axi-symmetric MHD waves are separated  in axi-symmetric  Alfv\'{e}n waves with $\xi_{\varphi} \neq 0, \; P'  =  0  $ and sausage magneto-sonic waves with $\xi_{\varphi} = 0, \; P'  \neq 0$. The pure sausage modes in a straight cylidrical flux-tube will be discussed in sub-section 3.2. For an axi-symmetric non-uniform 1-dimensional cylindrical plasma this is  the  only case where pure Alfv\'{e}n waves  show up in the analysis.  Each magnetic surface oscillates with its own local Alfv\'{e}n frequency.  In what follows we shall not be concerned with axi-symmetric Alfv\'{e}n waves.

Similary use  equations (53) of Goossens et al. (2019) for the components of   $\nabla \times \vec{\xi}$. 
Because of limitations of space  we only list the parallel component
\begin{eqnarray}
 (\nabla \times \xi)_z  & = & -  i \frac{m}{r} \frac{\displaystyle 1} {\displaystyle \left \{\rho_0 (\omega^2 - \omega_A^2) \right \} ^2} \frac{\displaystyle d}{\displaystyle dr} \left \{\rho_0 (\omega^2 - \omega_A^2)\right \}  P'
\label{EqStrFieldB}
\end{eqnarray}

\noindent 
The equations (\ref{EqStrFieldA}) and (\ref{EqStrFieldB})  clearly show that  $P'$ plays the role of coupling function (see also Hasagawa and Uberoi, 1982).  The horizontal components of vorticity $(\nabla \times \xi)_r $ and $(\nabla \times \vec{\xi})_{\varphi}$ are always non-zero. The parallel component $(\nabla \times \vec{\xi})_{\parallel} = (\nabla \times \xi)_z$, which can be considered as a marker for Alfv\'{e}n wave behaviour,  is non-zero when
\begin{equation}
\frac{\displaystyle d}{\displaystyle dr} \left \{\rho_0 (\omega^2 - \omega_A^2)\right \} \neq 0
\label{NonUniformOmegaA}
\end{equation}
Non-uniformity generates wave behaviour that is remeniscent of Alfv\'{e}n waves. In addition compression is non-zero and the wave has a mixed Alfv\'{e}n-magnetosonic behaviour.  MHD waves with frequencies in the Alfv\'{e}n continuum  undergo resonant Alfv\'{e}n  wave damping.  These waves have non-zero parallel vorticity in the non-uniform region and  parallel vorticity  becomes very big in absolute value when we move closer to the resonant position $r_A$ where $\omega_A(r_A) = \omega$.  For a resonantly damped wave the frequency is complex and there is not a singularity on the real axis but  parallel vorticity becomes very big anyway. Close to the ideal resonant position the Alfv\'{e}n behaviour dominates over the magnetosonic behaviour. The wave is an Alfv\'{e}nic wave.
%
%
%
Kink waves are non-axisymmetric waves with azimuthal wave number $m=1$. Hence the properties discussed in above 
apply to kink waves. Kink waves are important because they produce translational motions and are invoked to explain the transverse waves observed in flux tubes. Although this manifestation of kink wave is universally valid, however, if we consider such oscillations in the chromosphere, there are few scientific papers that report the observations of kink waves (Kukhianidze et al. 2006; Zaqarashvili et al. 2007; Morton et al. 2012; Kuridze et al. 2013; Morton et al. 2014; Stangalini et al. 2017; Jafarzadeh et al. 2017). A striking property of transverse MHD waves on magnetic tubes is their fast damping with damping times that are of the order of a few periods. It should be noted that this property is eventually an observational fact. Actually, the theory predicts that the damping time can take many oscillation periods too, if the profile of the perpendicular non-uniformity is steep.  A possible mechanism to explain the rapid damping of the transverse motions is resonant absorption (Hollweg and Yang 1988; Goossens et al. 1992; Ruderman and Roberts 2002; Goossens et al. 2002). Although, the below mentioned descriptions are universally valid for the radially and longitudinally structured magnetic flux tubes, however, here we analyze them in MHD regime in connection to the chromospheric tubes where the wave modes are evolved naturally.

Goossens et al. (2009) studied forces and analyzed  dispersion relations for several cases. In their section 3 they found that  kink waves do not disappear in incompressible MHD. This subject  was studied  in greater detail by Goossens et al. (2012). In subsection 4.2.2 of their paper, they studied kink wave ($m=1$) on a cylinder with piece-wise constant density. When they considered  the incompressible limit they found that all radial overtones disappeared but that the fundamental radial mode survived (See their figure 1). The fundamental radial mode of kink waves does not need compressibility. It is hard to call it a fast mode. It behaves very similar to surface Alfv\'{e}n waves in a Cartesian system as already emphasized by Wentzel et al. (1979). In the low beta limit the fundamental radial mode of kink waves has mixed properties. it is compressible as a magnetoacoustic wave but has also non-zero parallel vorticity as an Alfv\'en waves. This is explained in, e.g., Goossens et al. (2021). When the discontinous variation of density is replaced by a continuous variation in a transitional layer Goossens et al. (2012) found that  the fundamental radial mode of kink waves is resonantly damped but has both non-zero Eulerian perturbation of total pressure  everywhere and non-zero parallel vorticity in the non-uniform transitional layer. So it is a Alfv\'{e}nic surface wave with mixed properties. The ratio of parallel vorticity to compression depends on position. So the nature of the wave changes when it propagates through the plasma. Goossens et al. (2009) compared  the force  due to  magnetic tension  and the gradient total pressure force. They found that the magnetic tension force is the dominant force especially in the non-uniform layer. The fundamental radial mode of kink waves is clearly identified as an Alfv\'{e}nic wave that has both non-zero total pressure and non-zero parallel vorticity. The observation by Wentzel (1979) that surface waves differ from the  ordinary hydromagnetic waves (Alfv\'{e}n, slow, fast magnetosonic waves) is indeed correct. The energy content and propagation of kink MHD waves are investigated by Goossens et al. (2013a). Again  behaviour  of kink MHD waves differs substantially from that of bulk Afv\'{e}n waves and of magneto-acoustic waves.

As for the motion associated with the fundamental radial mode of kink waves we first consider the thin tube approximation of a piece-wise constant density flux tube. The Lagrangian displacement in horizonal planes inside the flux tube ($ 0 \leq r \leq R$) is (we suppress the dependence on $z$ and $t$; see Goossens et al. 2009, Goossens et al. 2014). 

\begin{equation}
\vec{\xi}_h(r, \varphi) =  \;C  \; (\cos \varphi \; \vec{1}_r - \; \sin \varphi \; \vec{1}_{\varphi}) \;\;
=  C \vec{1}_x = \; \vec{\xi}_{TR}
\label{TR}
\end{equation}
Equation (\ref{TR}) describes a uniform  motion of  the entire internal plasma  along the $x$-axis. The reason for this result is that $\xi_r$ and $\xi_{\varphi}$ have equal amplitudes and $\xi_{\varphi}$ is a quarter of period ahead of $\xi_r$. In the thin tube approximation the square of the frequency of the kink wave is
\begin{equation}
\omega^2 = \frac{\displaystyle \rho_i  \omega_{Ai}^2 + \rho_e
\omega_{Ae}^2} {\displaystyle \rho_i + \rho_e} = \omega_k^2.
\label{KinkFre}
\end{equation}
The density is piece wise constant and for that reason parallel vorticity is a $\delta$-function centered at the boundary explained in Goossens et al. (2012). The displacement field is shown in fig. 2 of Goossens et al. (2014).  Damping due to resonant absorption requires non-uniformity. When the discontinuous variation of equilibrium density is replaced by a continuous variaton of density from its internal value $\rho_i$ to its external value $\rho_e$ in a transitional layer of thickness $l$ then the eigenvalue of the kink wave is in the Alfv\'{e}n  continuum and the kink wave is resonantly damped (Goossens et al. 1992).  In addition, the non-uniformity generates parallel vorticity that is largest in absolute value at the resonant positiion. $\xi_r$ and $\xi_{\varphi}$ no longer have equal amplitudes The result is that the motion of the flux tube is a  translation combined with a rotational motion . The displacement field of a flux tube with a non-uniform transitional layer  is shown on fig.2 that is equivalent to fig. 11 of Goossens et al. (2014). In the vicinity of the resonant surface the motion resembles the motion depicted in figure 1 of Spruit (1981) for a non-axisymmetric $m=1$  Alfv\'{e}n wave in a uniform cylinder. Compression and the components of vorticity are shown in figures 1 and 2 of Goossens et al. (2020). Compression  is non-zero everywhere. Parallel vorticity is zero in the uniform part of the flux tube. It is non-zero in the non-uniform part and becomes large as we move closer to the resonant position. The temporal evolution  of a flux tube that is initally given a translational motion is shown in four snapshots in Fig. 12 of Goossens et al. (2014). The inital translational motion is transformed into a motion that is dominated by rotational motions. The dissipationless damping by resonant absorption is due to a transformation of energy of tanslational motions to energy of rotational motions as explained in Goossens et al. (2014).

\begin{figure}
\begin{center}
\includegraphics[scale=0.5]{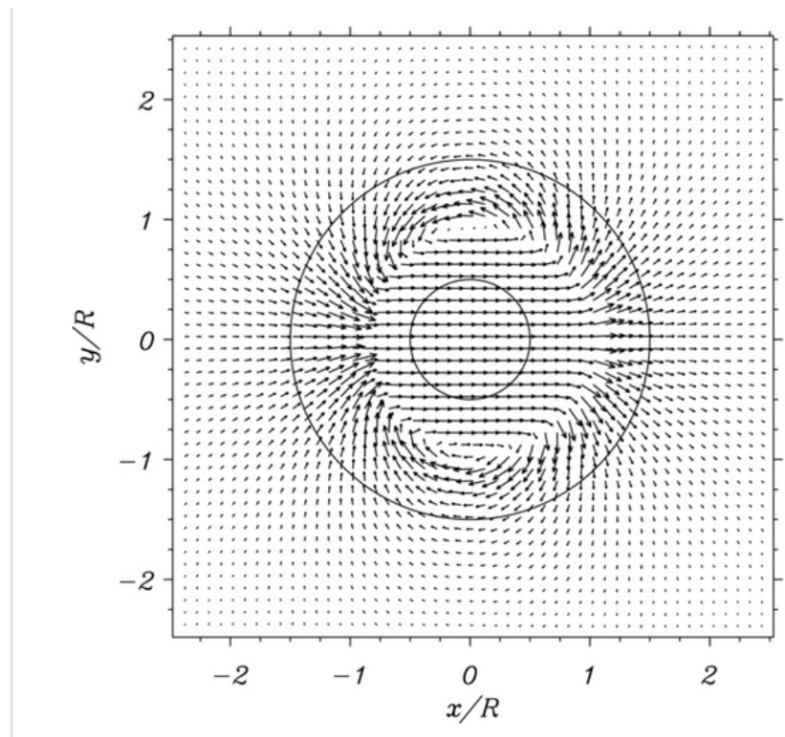}
\caption{The displacement field of a magnetized flux tube with a non-uniform transitional layer during the kink mode oscillations.
This figure is adapted from Goossens et al. (2014) (\copyright AAS \ Reproduced with permission).}
\end{center}
\end{figure}

%
%
\subsection{Sausage waves}
In sub-section~3.1, we have discussed the concept of all possible MHD modes and their properties in non-uniform and structured magnetic flux tubes. In the present sub-section, we discuss the physics and recent trends of sausage waves evolved individually in magnetic flux tubes in the solar chromosphere where we do not consider radial non-uniformity.
Sausage waves are axisymmetric modes in magnetic flux tubes with the azimuthal wave number $m=0$ (see Sect.~\ref{Sec:kink}). The waves modify the tube cross sections and hence density, therefore they have been frequently observed by radio observations in flaring coronal loops (Aschwanden et al. 2004), though additional efforts should be taken into account (Gruszecki et al. 2012; Reznikova et al. 2015). The sausage waves also modify plasma emission intensity, therefore can be revealed by imaging observations in the whole solar atmosphere, and also in solar chromospheric structures (Srivastava et al. 2008b; Morton et al. 2008; Grant et al. 2015).
 
\subsubsection{Linear sausage waves} 
 
Sausage waves in magnetic tubes are divided into fast and slow waves corresponding to the fast and the slow MHD modes (Edwin and Roberts 1983). Fast and slow sausage waves may have surface (with one velocity node at the tube center) and body (more than one velocity node along the radial direction) modes depending on the plasma parameters inside and outside the tube. In photospheric conditions, fast body sausage modes are absent (Edwin and Roberts 1983). The upper left panel of Figure 3 shows the dispersion diagram of fast and slow sausage waves under the conditions $v_A=2 c_s$, $v_{Ae}=0.5 c_s$ and $c_{se}=1.5 c_s$. Besides the phase speed, another important difference between the fast and slow sausage modes is the phase relation of the oscillations in cross sectional area and intensity. Fast surface sausage waves display anti-phase relation between the tube cross section and the emission intensity, while the slow surface and body waves show in-phase relation (Moreels et al. 2013). Therefore, the imaging observations may show the distinction between fast and slow modes. Another important feature of sausage waves is the cut-off wavenumber for fast surface sausage waves (Aschwanden et al. 2004)
\begin{equation}
k=k_c=\left [ \frac{(c^2_s+v^2_A)(v^2_{Ae}-v^2_T)}{(v^2_{Ae}-v^2_A)(v^2_{Ae}-c^2_s)} \right ]^{1/2}\frac{j_{0,s}}{R},
\end{equation}
where $v_T$ is the tube speed, $R$ is the tube radius and $j_{0,s}=(2.40, 5.52...)$ are the zeros of the Bessel function $J_0$. In photospheric conditions only the fast surface wave is allowed, therefore $j_{0,1}=2.40$.

The sausage waves may carry energy from the photosphere upwards into the chromosphere and corona contributing to the plasma heating processes. Therefore, the observation of waves at different levels of the solar atmosphere is of great importance. The first possible identification of linear sausage waves in photospheric magnetic pores was made by Dorotovi{\v{c}} et al. (2008) as oscillations of the cross sectional area of pores with periods of 20-70 min.\footnote{However such long periods require much longer vertical wave length than the width of the photosphere-chromosphere, therefore the interpretation of the observation in terms of sausage waves may be doubtful.} Morton et al. (2011) have found an observational evidence of high frequency sausage waves with periods
from 50 to 600 s. Fujimura and Tsuneta(2009) reported possible spectro-polarimetric observations of slow sausage waves in magnetic pores with Hinode. 

Grant et al. (2015) presented multi wavelength observations of magnetic pores using Dunn Solar Telescope at Sacramento  Peak, New Mexico. They used simultaneous observations of continuum, G-band, $Na \, I \,D_1$ and $Ca \, II \, K$ lines covering up to 800 km heights from the Sun's surface. Sausage mode oscillations with periods from 180 to 412 s, with an average period of 290 s, were detected in both intensity and area fluctuations. The phase difference between different wavelength bands showed upward propagation of sausage mode waves with a phase speed of $\sim$ 3 km s$^{-1}$. In-phase relation between oscillations in cross sectional area and intensity categorises the waves as slow sausage waves. The energy flux of the waves was estimated at different heights using the theoretical tool of Moreels et al. (2015a). They found the energy flux of the waves at the solar surface to be 35 kW m$^{-2}$, which rapidly decreased to 100 W m$^{-2}$ at the height of 800 km (see the upper right panel of Figure 3). The observed rapid reduction of the waves is not yet fully explained (Gilchrist-Millar et al. 2021; Riedl et al. 2021). Neither resonant absorption of slow sausage waves nor the electric resistivity is efficient enough to explain the observed damping (Yu et al. 2017; Chen et al. 2018). Ion-neutral collisions through Cowling resistivity may have a stronger effect, but it is generally much smaller in longitudinal slow waves than in transverse fast waves (Zaqarashvili et al. 2011a). Therefore, ion-neutral damping can probably not explain these observations.
Another possibility is slow mode conversion into fast/Alfv\'en modes in linear (Bogdan et al. 2003) and nonlinear (Ulmschneider et al. 1991; Zaqarashvili and Roberts 2006; Kuridze and Zaqarashvili 2008) regimes. However, estimates show that the  observed drop of energy flux occurred within one-quarter of the wavelength of the waves, therefore none of the damping/conversion mechanisms may explain the huge drop of the energy flux.
  
Observations reveal simultaneous propagation of sausage and kink waves in chromospheric structures, which may indicate their mutual coupling/conversion. Using the Dunn Solar Telescope, McAteer et al.(2003) observed propagating transverse kink (with frequency of 1.3, 1.9 mHz) and longitudinal sausage (with frequency of 2.6, 3.8 mHz) waves in the chromospheric network. These observations clearly showed the coupling of transverse and longitudinal oscillations with $\omega_l=2\omega_t$, where $\omega_t$ and $\omega_l$  are the transverse (kink) and the longitudinal (sausage) wave frequencies, respectively. This may indicate nonlinear coupling of transverse and longitudinal waves, which satisfy the theoretical frequency relation of coupled waves (Ulmschneider et al. 1991; Zaqarashvili and Roberts 2006). Morton et al.(2012) detected clear coupling of kink and sausage waves in chromospheric structures (see lower panel of Figure 3). Clear anti-phase relation between oscillations of structure width and emission intensity ranked the sausage waves as fast waves. The observed periods of kink and sausage waves were estimated as 232 $\pm$ 8 s and 197 $\pm$ 8 s, which are clearly out of nonlinear resonant conditions, probably indicating linear coupling of the waves.
  
Dorotovi{\v{c}} et al.(2014) detected fast and slow sausage waves with periods from 4 to 65 min in photospheric magnetic waveguides using the Swedish 1-m Solar Telescope. The period ratio of the oscillations indicated that they are part of a group of standing harmonics in a flux tube that is non-homogeneous and bound by the photosphere and the transition region. Later Freij et al.(2016) confirmed the existence of standing harmonics of slow sausage waves, which enabled the estimation of tube expansion factor that was in good agreement with numerical simulations (see also Moreels et al. 2015b).
  
Keys et al.(2018) presented direct evidence of surface and body sausage waves in numerous magnetic pores at the solar photosphere. The authors found surface modes more frequently than body modes in the data. Observed wave frequencies were in the range of $\sim$ 2-12 mHz, where the  body mode frequency reached up to 11 mHz, but no surface modes were found above 10 mHz. The authors estimated that 35 kW m$^{-2}$ at the photospheric level the surface modes transport at least twice the average energy flux (22$\pm$10 kW m$^{-2}$) as the observed body modes (11$\pm$5 kW m$^{-2}$). This may be significant in determining which mode contributes more to localized atmospheric heating as a function of waveguide height.

Gafeira et al. (2017) reported the detection of high-frequency oscillations in slender Ca II H fibrils from high-resolution observations acquired
with the SUNRISE balloon-borne solar observatory. The fibrils show obvious predominantly anti-phase oscillations in their intensity and width, which classifies the waves as fast sausage modes. The obtained distributions have median values of the period of 32$\pm$17 s and 36$\pm$25 s, respectively.

As seen above, the recent high-resolution observations detect the presence of sausage waves in the chromosphere that are carrying substantial energy flux to heat it and the overlying corona. Another exclusive feature, which is rarely observed, but could be an another important aspect of the hydrodynamic response of solar chromosphere, is the slow sausage solitons. We describe it briefly in the next sub-section.
\subsubsection{Slow sausage solitons}
When slow sausage pulses propagate from the photosphere upwards they may quickly steepen into shocks due to the rapid decrease of density. In certain conditions, the tube dispersive effects may prevent the non-linear steepening leading to the formation of a soliton, which is a stable structure propagating without significant change of shape. The formation of sausage solitons in magnetic tubes was first suggested by Roberts and Mangeney(1982), which was followed by several papers about weakly nonlinear waves and solitons (Zhugzhda and Nakariakov 1997; Nakariakov and Roberts 1999; Zhugzhda 2005; Erd{\'e}lyi and Fedun 2006; Barbulescu  and Erd{\'e}lyi 2016). The solution of a slow sausage surface soliton in magnetic slabs with width $2d$
is given by the following expression (Ruderman 2003).
\begin{equation}
\eta=\frac{al^2}{l^2+ \mid z-st \mid }
\end{equation}
where $\eta$ is the displacement of the slab boundary, $a$ is the soliton amplitude and 
\begin{equation}
s=v_T +\frac{1}{4}\frac{a b}{d}, \,\, l=4\frac{\kappa d}{ab}.
\end{equation}
are the soliton speed and the spatial scale, respectively. The parameters $b$ and $\kappa$ are expressed as
\begin{equation}
b=\frac{v^4_A[3c^2_s+(\gamma+1)v^2_A]}{2v_T(v^2_A+c^2_s)^2}, \,\,\kappa=\frac{d}{2}\frac{\rho_e}{\rho_i}\frac{v_Tc^2_s(v^2_T-v^2_{Ae})}{m_ev^2_A(c^2_s+v^2_A)}.
\end{equation}

The typical symbols of characteristic speeds, densities, etc have their usual meanings as we have followed throughout in this review article ($v_T=v_Ac_s/(v_A+c_s)$ is the tube speed). Barbulescu  and Erd{\'e}lyi (2016) studied the nonlinear sausage waves in magnetic flux tubes and concluded that the slow sausage soliton can be formed in certain conditions.

Zaqarashvili et al.(2010a) analysed a time series of the Ca II H line obtained at the solar limb with the Solar Optical Telescope (SOT) on board Hinode (see also Zaqarashvili et al. 2011b). Observations showed an intensity blob  propagating from 500 to 1700 km
above the solar surface with a mean apparent speed of 35 km s$^{-1}$. The speed was much higher
than the expected local sound speed, therefore the blob could not be a simple pressure pulse. The authors found that the
blob speed, length-to-width ratio and relative intensity were characteristic of a slow sausage soliton
propagating along a magnetic tube. The blob width increased with height corresponding to the
magnetic tube expansion in a stratified atmosphere. The authors suggested that the propagation of the intensity blob may
be the first observational evidence of a slow sausage soliton in the solar atmosphere. 

 \begin{figure}
\noindent\includegraphics[width=6cm]{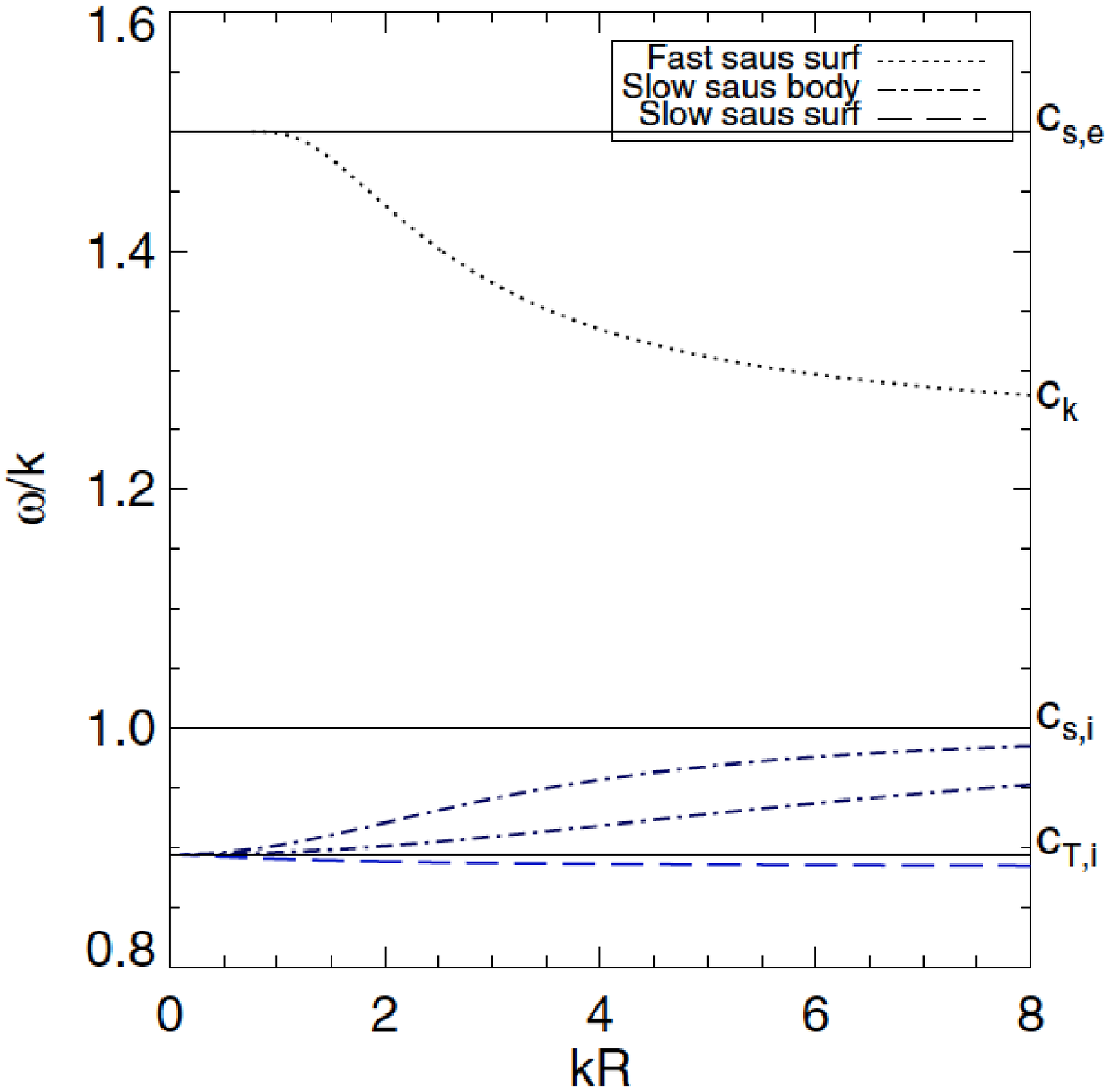}
\noindent\includegraphics[width=8cm]{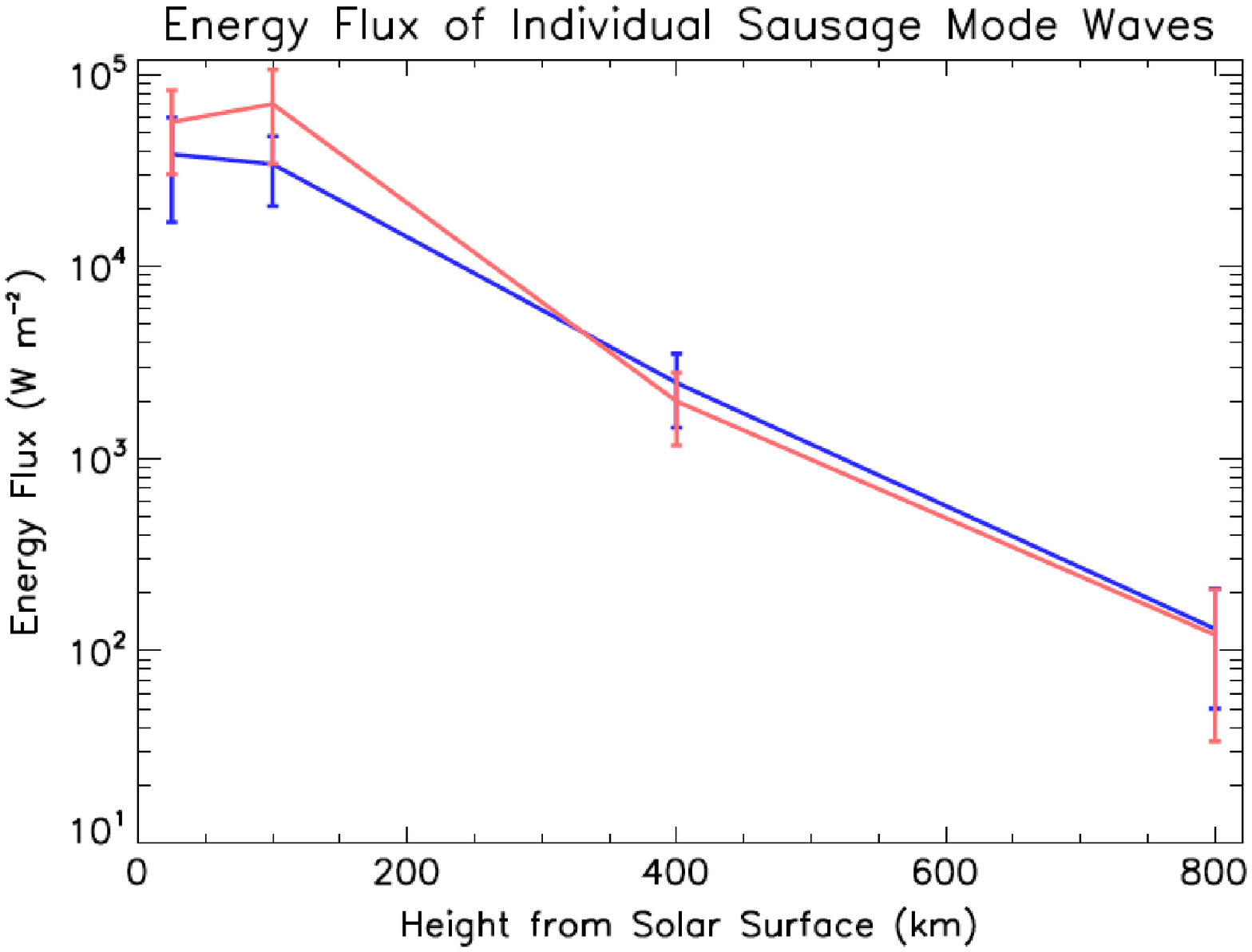}
\noindent\includegraphics[width=\textwidth]{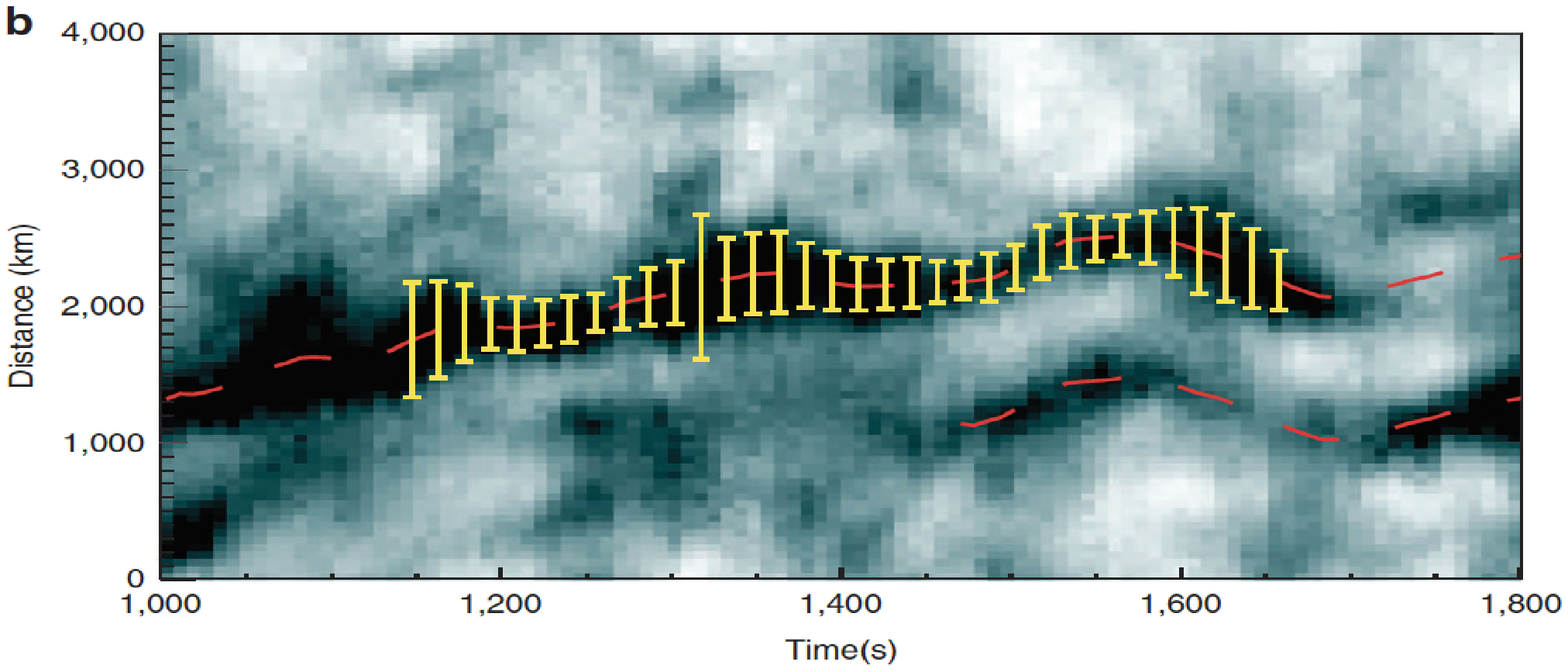}
\caption{Upper left panel: Phase speed diagram of sausage waves under photospheric conditions: $v_A=2 v_s$, $v_{Ae}=0.5 v_s$ and $v_{se}=1.5 v_s$. Fast body waves are absent in these conditions. This panel is adapted from Moreels et al.(2013; $\copyright$ ESO \ Reproduced with permission). Upper right panel: The calculated energy flux of 210 s (red) and 290 s (blue) sausage mode oscillations vs height from the solar surface observed by the Dunn Solar Telescope. Rapid decrease of energy flux of both oscillations with height are clearly seen. This figure panel is adapted from Grant et al. (2015; $\copyright$ AAS \ Reproduced with permission). Lower panel: Time-distance plot from ROSA observations showing simultaneous propagation of kink and sausage modes in chromospheric structures. Transverse displacement corresponds to the kink wave and the variation of structure width corresponds to the sausage wave. This panel is adapted from Morton et al.(2012).} 
\label{Figure-tz}
\end{figure}

In the next section, we emphasize the basic physics of torsional modes in the solar chromosphere, and the recent trend of scientific research keeping the view of solar chromosphere at the central place.


\subsection{Torsional waves}
\label{sec:torsional_waves}
\noindent The presence of Alfv\'en waves in the Earth's magnetosphere is unquestionable, with their role in magnetosphere--ionosphere coupling, energy transportation, field line resonance, and particle acceleration at the forefront of solar terrestrial physics (Keiling 2009). Contrarily, Alfv\'en waves have long been one of the most elusive waveforms in the solar atmosphere (Mathioudakis et al. 2013). However, the desire to identify and benchmark the capabilities of Alfv\'en waves manifesting throughout the Sun's atmosphere stems from the original pioneering work by Alfv{\'e}n (1942), who put forward the idea that these waveforms may be responsible for the elevated temperatures found in the solar corona (Alfv{\'e}n 1947). 

Such speculation is a result of the intrinsic properties of Alfv\'en waves, whereby their relative incompressibility (when compared to other MHD waves, including fast and slow magnetoacoustic modes) allows them to propagate much further before being dissipated. The incompressibility of Alfv\'en waves is the result of magnetic tension providing the only restoring force when driven by linear perturbations. Hence, to achieve dissipation of the energy embodied in Alfv\'en waves in order to provide thermal energy to the outer solar atmosphere, specific mechanisms must be invoked, including phase mixing (Heyvaerts and Priest 1983; Ofman and Aschwanden 2002; Ebadi et al. 2012; Prokopyszyn and Hood 2019; Van Damme et al. 2020), resonant absorption (Ionson 1978; Davila 1987; Poedts et al. 1989, 1990; Ofman et al. 1994,1995; Goossens et al. 2006, 2011; Giagkiozis et al. 2016; Howson et al. 2019), mode conversion (Crouch and Cally 2005; Suzuki and Inutsuka 2005; Cally and Khomenko 2015; Pagano and De Moortel 2017), and Alfv\'en turbulence (Cranmer and van Ballegooijen 2005; van Ballegooijen et al. 2011,2017; van Ballegooijen and Asgari-Targhi 2016; Oran et al. 2017; Dinesh Singh and Singh Jatav 2019). Indeed, Chitta et al.(2012) compared the velocity fluctuations corresponding to small-scale magnetic elements in the lower solar atmosphere and found significant Fourier power associated with high-frequency ($\sim20$~mHz) horizontal motions, hence providing indirect evidence for the creation of a turbulent environment that can efficiently provide Alfv\'en wave dissipation. Furthermore, it is also possible for linear Alfv\'en waves to dissipate their initial energy through the process of parametric decay, which combines a weakly turbulent environment with the coupling of Alfv\'en waves to other compressible magnetoacoustic modes (Malara and Velli 1996). In this regime, the initial energy of the Alfv\'en wave is gradually transferred to daughter species, resulting in the destruction of the initially coherent state. This form of energy transfer has been identified in the solar wind (Malara et al. 2000; Del Zanna 2001; Tenerani and Velli 2013; Primavera et al. 2019), but has yet to be fully documented in the confines of the solar atmosphere.

In sub-section~3.1, we have discussed the physics of all possible MHD modes (kink, sausage, and Alfv\'enic) and their properties in non-uniform and structured magnetic flux tubes. In the present sub-section, we describe only physics and recent research on pure incompressible torsional Alfv\'en waves that are generated individually in the magnetic flux tubes (over iso-frequency magnetic surfaces) where no radial non-uniformities exist. Here, we use the term `radial' to indicate the direction perpendicular to the magnetic field vector. We essentially exclude the concept of Alfv\'enic waves (mixed mode of radial kink waves with surface Alfv\'en waves) in such physical conditions, and the pure Alfv\'en waves are likely evolved in the magnetized solar chromosphere. In such a situation, once the driving forces that underpin the Alfv\'en waves become sufficiently non-linear, magnetic tension still remains the dominant restoring force, but the validity of incompressibility may become compromised. For circularly polarised Alfv\'en waves, they remain incompressible due to the strict rotation of their magnetic field vector around the direction of propagation (Ferraro 1955). However, linearly or elliptically polarised non-linear Alfv\'en waves create density perturbations in the local plasma due to the operation of ponderomotive forces, with the resulting dynamics governed by the derivative of the non-linear Schr{\"{o}}dinger equation (Medvedev 1999; Laming 2009),
\begin{equation}
\label{eqn:derivativeSchrodinger}
{{\partial b}\over {\partial \tau}}+ {{1}\over {4(1-\beta)}} {{\partial }\over {\partial z}}(|b|^2b)\pm i {{v^2_A}\over {2\Omega_i}} {{\partial^2 b}\over {\partial z^2}}=0 \ ,
\end{equation}
where $b=B_{\perp}/B_z$ is the relative amplitude of transverse magnetic field perturbations along a straight unperturbed magnetic field, $B_{z}$, $\Omega_i$ is the ion cyclotron frequency, $\tau=(B_z/B_{\perp})^2t$ is the extended time, and $\beta\propto{c}^2_s/v^2_A$ is the ratio of the plasma pressure to the magnetic pressure, where $v_A$ and $c_s$ are the local Alfv\'en and sound speeds, respectively. Within the lower solar atmosphere, the ion-cyclotron frequency has values on the order of $10^{5} - 16^{6}$~Hz (Khomenko et al. 2014b), which helps to alleviate wave dispersion and simplifies Equation~{\ref{eqn:derivativeSchrodinger}} to,
\begin{equation}
\label{eqn:simplifiedSchrodinger}
{{\partial b}\over {\partial \tau}}+ {{1}\over {4(1-\beta)}} {{\partial }\over {\partial z}}(|b|^2b)=0 \ .
\end{equation}
Hence, non-linear Alfv\'en waves are able to steepen into shocks (Cohen and Kulsrud 1974; Murawski et al. 2015; Shestov et al. 2017; Snow et al. 2018), with such steepening being expedited when the Alfv\'en speed is approximately equal to the local sound speed (i.e., the plasma-${\beta\sim1}$; Montgomery 1959).
 Of course, it is important at this stage to highlight the differences between the non-linear evolution of plane shear Alfv\'en waves and torsional oscillations in magnetic flux tubes. For both cases, Vasheghani Farahani et al. (2012) found that a non-linear self-interaction resulted in the steepening of the waves. However, in the case of a finite plasma-$\beta$, the non-linear steepening of torsional Alfv\'en waves develops more slowly than their plane shear wave counterparts. Interestingly, Vasheghani Farahani et al. (2011) found that the precise value of the plasma-$\beta$ had no effect on the creation of compressive perturbations in propagating torsional Alfv\'en waves, something that was linked to the inherent tube speed (i.e., the natural speed of the longitudinal compressive perturbations) always being lower than both the local Alfv\'en and sound speeds. Recent evidence for such Alfv\'en wave steepening has been put forward by Grant et al. (2018), who employed a combination of imaging spectroscopy and inversion routines to examine the temperature fluctuations of chromospheric plasma when Alfv\'en wave signals were present. 

However, conclusively detecting the presence of Alfv\'en waves is a challenging endevour. It is common to detect the appearance of MHD wave modes (e.g., the slow magnetoacoustic mode) through the intensity fluctuations generated by the compressions and rarefactions of the propagating wave signal. Unfortunately, the lack of compressibility for linear and circularly polarised Alfv\'en waves means that there are no associated periodic signatures in the intensity signal being emitted by the wavetrain. Thankfully, the lower solar atmosphere provides structures and driving mechanisms that should result in an abundance of Alfv\'en waves. For example, the most basic form of Alfv\'en waves, the torsional mode with an azimuthal wavenumber $m=0$, can be readily generated in axially symmetric structures synonymous with sunspots, magnetic pores, spicules, and filaments through the ubiquitous buffeting motions present at the footpoints of these omnipresent features (Ruderman et al. 1997; Ruderman 1999; Luo et al. 2002; Matsumoto and Shibata 2010; Fedun et al. 2011a; Vigeesh et al. 2012). Indeed, Shelyag and Przybylski (2014) employed simulated photospheric Fe~{\sc{i}} 6302~{\AA} observations generated by the MURaM (V{\"o}gler et al. 2005) MHD code to search for the presence of Alfv\'en waves at the base of the photosphere. The authors found an abundance of evidence for Alfv\'en waves manifesting in magnetic elements in the intergranular lanes, however, these signatures disappeared when the simulated observations were degraded to the resolution of the Solar Optical Telescope (SOT; Tsuneta et al. 2008)
 onboard the Hinode (Kosugi et al. 2007) spacecraft, hence highlighting the need for high resolution observations of the solar photosphere and chromosphere to study Alfv\'en wave generation and propagation. 

\begin{figure}[!t]
\noindent\includegraphics[width=\textwidth]{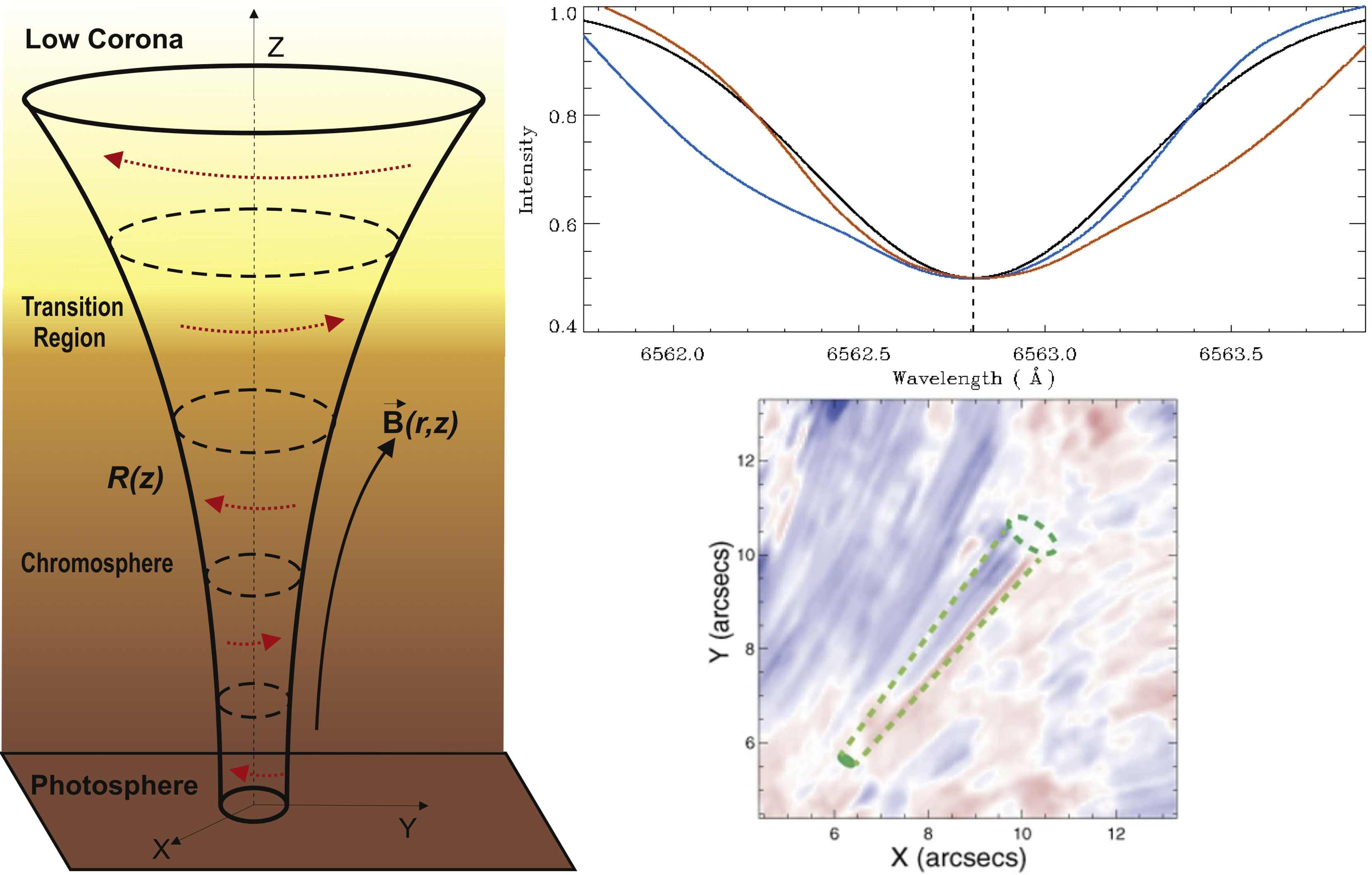}
\caption{{\textit{Left:}} A cartoon schematic of Alfv\'en wave propagation in an expanding waveguide, where the rotation of magnetic isosurfaces synonymous with the presence of torsional Alfv\'en waves is highlighted using dotted red arrows (image reproduced from Soler et al. 2017; $\copyright$ AAS \ Reproduced with permission). {\textit{Upper-right:}} Sample H$\alpha$ spectra corresponding to heavily blue-wing broadened (blue line), quiescent (black line), and heavily red-wing broadened (red line) profiles that demonstrate the non-thermal broadening characteristics consistent with the presence of torsional Alfv\'en waves. Here, the blue, black, and red H$\alpha$ profiles were acquired simultaneously by the SOUP instrument at the SST and correspond to the spectra obtained at the center (black line) and opposite edges (blue and red lines) of the chromospheric waveguide (image adapted from Jess et al. 2009, Mathioudakis et al. 2013; $\copyright$ AAS \ Reproduced with permission). {\textit{Lower-right:}} Line-of-sight H$\alpha$ Doppler velocities acquired by the CRISP instrument at the SST, where the dashed green lines outline the boundaries of a chromospheric magnetic flux tube and the axes represent heliocentric coordinates. The asymmetric red (down flowing) and blue (up flowing) velocities across the diameter of the waveguide indicate the presence of torsional Alfv\'en waves embedded within the magnetic structure (image adapted from Srivastava et al. 2017). }
\label{fig:torsional_JESS}
\end{figure}

With cutting-edge simulations verifying the mechanisms responsible for generating ubiquitous Alfv\'en waves in the lower solar atmosphere, attention naturally turns to the identification of these features in modern ground- and space-based observations. As highlighted above, the relative incompressibility of Alfv\'en modes means that intensity fluctuations cannot be harnessed to provide evidence of such waves. Instead, attention needs to be turned to spectroscopy. An $m=0$ torsional Alfv\'en wave propagating along a magnetic flux tube can be visualised as a set of rotating magnetic iso-surfaces (left panel of Figure~{\ref{fig:torsional_JESS}}). When the waveguide is viewed from the side, the rotation of the iso-surfaces manifest in equal, yet opposite Doppler shifts due to one side of the flux tube rotating towards the observer, while the other side of the flux tube rotates away from the observer. If the Alfv\'en waveguide is well resolved spatially, then the observer will be able to detect the induced blue- and red-shifted Doppler velocities manifesting at opposite edges of the flux tube (Mathioudakis et al. 2013; Srivastava et al. 2017; Kohutova et al. 2020). However, if the Alfv\'en waveguide is below the resolution element of the telescope instrumentation, then the opposing Doppler shifts will become blended and result in periodic fluctuations in the full-width at half-maximum (FWHM) of the measured spectral line (McClements et al. 1991; Zaqarashvili 2003). The magnitude of such non-thermal line broadening will be dependent on both the velocity amplitude of the Alfv\'en wave, as well as the inclination angle of the waveguide with respect to the observer's line-of-sight (i.e., the $\cos$ term will modulate the degree of Doppler-induced broadening visible to the observer). Furthermore, turbulence and/or bulk flows embedded within the flux tube may also mask the presence of FWHM oscillations caused by Alfv\'en waves (Jess and Verth 2016). As a result, a combination of high spatial resolution observations and high precision spectroscopy is required to unequivocally detect the presence of Alfv\'en waves manifesting in the lower solar atmosphere. 

Employing Solar Optical Universal Polarimeter (SOUP; Title 1984)
observations of the chromospheric H$\alpha$ absorption line, Jess et al. (2009) provided the first detection of Alfv\'en waves in the solar chromosphere. Through examination of a magnetic flux tube anchored above a photospheric magnetic bright point, Jess et al. (2009) measured the amplitude of H$\alpha$ non-thermal line broadening to be $\sim0.05$~{\AA} (or a velocity amplitude on the order of $2.5$~km/s; upper-right panel of Figure~{\ref{fig:torsional_JESS}}), and when combined with a plasma density of approximately $10^{-9}$~g/cm$^{3}$ and a local Alfv\'en speed of $\sim22$~km/s, provided a wave energy flux equal to $\sim150{\,}000$~W/m$^{2}$. While this quantity of energy is far in excess of the amount required to balance chromospheric radiative losses (Withbroe and Noyes 1977), Verth and Jess (2016) hypothesised that the filling factor of such waves needs to be evaluated before the true importance of their existence can be evaluated. With this in mind, Fedun et al. (2011b) harnessed the numerical Sheffield Advanced Code (SAC; Shelyag et al. 2008) and placed a vortex-type driver at the photospheric boundary, before examining the torsional effects and higher atmospheric heights. Comparing the outputs from the SAC code with the observational findings of Jess et al. (2009), Fedun et al. (2011b) revealed how the geometry of magnetic flux concentrations that make up realistic solar structures (e.g., magnetic bright points, pores, etc.) acts as a spatial frequency filter that modulates the observable wave signatures, hence providing a novel method to map solar magnetism as a function of atmospheric height by capturing torsional Alfv\'en waves in different spectral lines with corresponding differences in their formation heights.

While the work of Jess et al. (2009) examined Alfv\'en waves present in long-lived magnetic flux tubes extending upwards from the photosphere, De Pontieu et al. (2012) turned attention towards much more dynamic and rapidly evolving chromospheric features in the form of Type-{\sc{ii}} spicules (Langangen et al. 2007; Rouppe van der Voort et al. 2009; Kuridze et al. 2015). Harnessing a combination of both the Crisp Imaging Spectropolarimeter (CRISP; Scharmer et al. 2008) and the TRI-Port Polarimetric Echelle-Littrow (TRIPPEL; Kiselman et al. 2011) spectrograph on the Swedish 1-m Solar Telescope (SST), De Pontieu et al. (2012) found evidence that the dominant majority of Type-{\sc{ii}} spicules undergo large torsional twists that represent the signatures of Alfv{\'{e}}nic waves propagating towards the corona at several hundred km s$^{-1}$. This interpretation was reached due to the ability of the CRISP and TRIPPEL instruments to resolve the red-blue Doppler velocity asymmetries across the diameter of the Type-{\sc{ii}} spicules in both Ca~{\sc{ii}}~H and H$\alpha$ observations. In this work, the term `Alfv{\'{e}}nic' is used to categorise the embedded wave motion since Type-{\sc{ii}} spicules also demonstrate large-amplitude transverse (kink) waves in addition to their torsional (with respect to the axis of the magnetic field) motions. Goossens et al. (2009, 2012) revealed that plasma compression is limited in the case of thin magnetic flux tubes (such as for Type-{\sc{ii}} spicules), and hence the resulting wave motion is more closely aligned with {\Alfven} waves than typical fast magneto-sonic waves. Hence, the term `Alfv{\'{e}}nic' was coined (Goossens et al. 2009) to impart a caveat that pure theoretical Alfv\'en waves (as described by Alfv{\'e}n 1942) can only exist in a uniform plasma of infinite extent. Goossens et al. (2014) also highlighted that the derived velocity profiles of transverse kink waves can lead to similar Doppler characteristics that would be expected for torsional Alfv\'en waves, and therefore the viewing angle of the solar structures must be taken into consideration to reach a firm conclusion regarding the captured wave mode. The details of the key theoretical developments on Alfv\'enic waves originally proposed by M. Goossens and fellow scientists over the last two decades are summarized in sub-section 3.1.

Following the launch of the Interface Region Imaging Spectrograph (IRIS; De Pontieu et al. 2014a), De Pontieu et al.(2014b) examined high spatial resolution ($\approx0.33$~arcsec) UV spectra and found that the chromosphere and transition region was replete with small-scale torsional motions. Importantly, De Pontieu et al.(2014b) found that such torsional motion was readily present in quiet Sun, active region, and coronal hole locations, highlighting the omnipresent nature of torsional behaviour in the solar chromosphere. Utilising the multi-thermal capabilities of the IRIS instrument, it was found that the chromospheric structures undergoing torsional motions were rapidly heated to transition region temperatures, which was hypothesised to be compatible with the heating expected from the dissipation of torsional Alfv\'en waves generated by small-scale photospheric vortices (van Ballegooijen et al. 2011; Asgari-Targhi and van Ballegooijen 2012; Shelyag and Przybylski 2014). If the chromosphere is indeed permeated by torsional Alfv\'en waves, then they are required to propagate through the temperature minimum region. Here, ion-neutral effects are likely to play an important role (Khomenko and Collados 2012; Khomenko et al. 2018), with theoretical work revealing that short period ($<5$~s) Alfv\'en waves damp very rapidly in the chromospheric network as a result of ion-neutral collisions (Zaqarashvili et al. 2013). Hence, in order to probe the heating effects of torsional Alfv\'en waves, the community is turning its attention towards the highest frequencies currently achievable with modern instrumentation. 

Recently, Srivastava et al. (2017) utilised the high-precision CRISP instrument on the SST to search for high-frequency torsional Alfv\'en waves. Nine wavelength steps across the H$\alpha$ absorption line were chosen to maximise the spectral cadence ($\approx$3.9~s), with the resulting images reconstructed using the multi-object multi-frame blind deconvolution (MOMFBD; van Noort et al. 2005) code to ensure the smallest spatial scales were visible. Following analyses using Fourier cross-correlation and wavelet techniques, Srivastava et al. (2017) uncovered evidence for the presence of torsional Alfv\'en waves with frequencies in the range of $12-42$~mHz, providing periodicities ($24-83$~s) approximately one order-of-magnitude more rapid than previously uncovered (Jess et al. 2009). The high spatial resolution of the CRISP instrument allowed the blue and red Doppler shifts associated with the rotating magnetic iso-surfaces to be examined (see the lower-right panel of Figure~{\ref{fig:torsional_JESS}}). Numerical models, including the use of the FLASH code (Lee and Deane 2009), revealed that the necessary high-frequency photospheric wave drivers were able to generate and supply vast Alfv\'en wave energy flux ($\sim 10^{5}$~W/m$^{2}$) into the solar chromosphere, consistent with previous analytical estimations (Murawski et al. 2015). Importantly, Srivastava et al.(2017) concluded that even accounting for partial reflections at the transition region interface, more than $\sim 10^{3}$~W/m$^{2}$ would be transmitted into the overlying corona. This has significant implications for studies into both the heating of the multi-million degree corona and the generation of the supersonic solar wind. 

Over the last decade, torsional Alfv\'en waves manifesting in the lower solar atmosphere have gone from elusive through to ubiquitously detectable in modern observing sequences (e.g., Jess et al. 2009; Srivastava et al. 2017; Grant et al. 2018; Liu et al. 2019). This success has arisen from a combination of more sensitive instrumentation, better wave detection techniques, alongside more refined MHD wave theory. Looking towards the future, we have already been stepping into an era of ultra-high precision solar physics with the advent of the 4-m Daniel K. Inouye Solar Telescope (DKIST; Rimmele et al. 2020) that has already seen its first light in 2020 and now releasing unprecedented fine details of the solar atmosphere. It will now become possible to examine the Doppler velocity asymmetries synonymous with torsional Alfv\'en waves on unprecedented spatial scales as small as $\approx$20~km (Rast et al. 2020). A current challenge is understanding the precise observables we will be able to detect at these small spatial scales. For example, when the magnetic flux tube supporting a torsional Alfv\'en wave is non-uniform (e.g., there are non-uniformities in the radial direction away from the central axis of the magnetic field), then elements of phase mixing will need to be considered (e.g., Heyvaerts and Priest 1983; Browning and Priest 1984; Shestov et al. 2017; Diaz-Suarez and Soler 2021). Here, the variable phase velocities of the embedded Alfv\'en waves cause refraction and hence create large gradients in the direction perpendicular to the magnetic field (Ruderman et al. 1999). When structures with reduced opacities (e.g., coronal loops) are examined, this naturally results in mixed observational signatures that may be difficult to disentangle. However, this concern may be less important in the optically thick chromosphere, where observations of fibrillar-type features more closely align with the outermost shells of the structure, hence providing less mixed signals due to larger local opacities (Bose et al. 2019). Nevertheless, phase mixing is able to produce Kelvin-Helmholtz instabilities in the observed plasma (Guo et al. 2019), which may need to be taken into consideration for accurate scientific interpretations. High resolution observations, combined with next-generation instrumentation capable of high cadence data acquisition, will make it possible to probe the ion-neutral effects of Alfv\'en wave dissipation at periodicities less than $5$~seconds. As a result, the community looks forward to making rapid advancements in the understanding of torsional Alfv\'en waves and their role in supplying energy at high frequencies to the outer regions of the solar atmosphere. 

In Section 3, we have discussed the physical properties of various MHD modes and their possible contributions in heating the solar chromosphere. In the next section we will discuss the overarching scenario of chromospheric heating, keeping a central view on MHD waves.

\section{Heating of the Chromosphere}\label{Sec:heating}
It Section.~\ref{Sec:structured}, it is noticed that MHD waves carry substantial energy in the solar chromosphere, therefore, in the context of the wider physical implications of the heating of this layer the role of these waves should be explicitly studied and explored. In this Section~\ref{Sec:heating}, we evaluate the heating aspects of the MHD waves in the solar chromosphere. In sub-section~\ref{Sec:heating_gen_ls}, we discuss in general the heating aspects of the large-scale chromosphere, while in the sub-section~\ref{Sec:heating_loc_tube} we will depict the energy budget of MHD waves in chromospheric localized flux tubes and thereby their heating capabilities. In the sub-section~\ref{Sec:heating_prom} we describe the wave heating of solar prominence plasma.
%
%
\subsection{Wave heating of the large-scale chromosphere}\label{Sec:heating_gen_ls}
No generally accepted theory of the formation of the hot solar chromosphere exists so far. The total radiative losses of the quiet chromosphere are of the order of 4300 W m$^{-1}$ (Avrett 1981) and up to a factor of 2-4 higher in active regions (Withbroe and Noyes 1977). Heating by waves falls into one of the most popular categories of chromospheric heating models, and it can be sub-divided into the heating by acoustic waves (without involving the magnetic fields) and heating by different kinds of {MHD waves} (Sections 2 and 3). The latter involves interactions of acoustic waves generated in sub-photospheric layers with magnetic structures present at the surface. In order to understand wave heating mechanisms in the chromosphere one needs first to understand which processes can help bringing the wave energy in sufficient amount to these layers and then to provide efficient dissipation mechanisms for these waves. Wave energy transport and dissipation are frequency-dependent, and their efficiency is different for high- and low-frequency waves. The maxima of the observed solar wave spectra fall into the 3-5 mHz range, depending on the height. This low-frequency part of the solar photospheric spectrum has been extensively studied both theoretically and observationally (Gizon and Birch 2005). The high-frequency part of the spectrum still represents a challenge from an observational point of view, while theoretically there is increasing evidence that high-frequency wave phenomena may play a relevant role in energizing the solar chromosphere. 

Heating of the quiet chromosphere can be done by acoustic waves (see Section 2). These waves are continuously produced by convection (e.g., Goldreich and Keeley 1977; Balmforth 1992; Nordlund and Stein 2001; Stein and Nordlund 2001). The amount of energy contained in acoustic waves is sufficiently high to heat the upper chromosphere and corona if all this energy could reach these high layers and be converted into heat there (e.g., Biermann 1946; Schwarzschild 1948; Stein 1968; Narain and Ulmschneider 1996). The change of the atmospheric properties in the vertical direction in a gravitationally stratified solar atmosphere leads to a steepening of the vertically propagating acoustic waves into shocks. The energy of the propagating shock waves is then dissipated near their fronts due to radiation or viscosity, converting it into the thermal energy. Such an energy deposit process has a stochastic character due to the stochastic nature of the excited acoustic waves (e.g., Ulmschneider 1971a,b; Ulmschneider et al. 1978; Carlsson and Stein 1997). According to Carlsson and Stein (1995, 1997), who modeled time-dependent acoustic waves propagation including non-LTE radiative losses in the chromosphere, the chromosphere is not necessarily hot at all times and locations. In their simulations, there are observed short intervals of very high temperature caused by acoustic shocks but the average chromospheric temperature continued to decrease with height. The heights where the shock formation and dissipation happen depend on the wave frequency and amplitude. Higher-frequency acoustic waves (periods around 10$^1$ s) shock rather low in the atmosphere (Ulmschneider 1971a,b; Narain and Ulmschneider 1996), and therefore must not be the best candidates for heating the upper atmosphere. Lower frequency waves (periods of 10$^2$ s) could propagate the energy higher up, but this process is limited by the presence of the acoustic cut-off frequency around 3 mHz at the bottom of the  photosphere. Thus, acoustic  waves with maximum energy in the spectrum are unable to transport this energy to the chromosphere. 

On their way to the upper atmosphere, the acoustic waves suffer multiple changes of their physics, related to the presence of magnetic structures in these layers. The magnetic field concentrations embedded in solar granulation suffer foot point motions and this drives waves in these structures through the solar atmosphere, as was shown in numerous ``idealized'' simulations of the wave dynamics (see Khomenko and Calvo Santamaria 2013; Khomenko and Collados 2015).  As the waves propagate upwards they find obstacles, such as the equipartition layer where the acoustic, $c_S$ and the Alfv\'en, $v_A$, speeds coincide $c_s=v_A$ (equivalent to plasma $\beta=1$), the layer where the local acoustic cut-off frequency is equal to the wave frequency, the steep temperature gradient at the transition region, etc. In these regions waves suffer transformations, refraction, and reflection. Some of these processes prevent the wave energy reaching the upper chromosphere and corona, but others can favor wave energy transmission. It is important to take into account the location of these critical layers with respect to the height of formation of the observed spectral lines, see e.g. Moretti et al. (2007). The efficiency of the wave-related heating mechanisms will depend on the structure and topology of the magnetic fields in a given solar region. 


Vast theoretical effort has been dedicated to understanding how low-frequency waves (3-5 mHz) can reach the chromosphere and corona with the help of magnetic field concentrations, and what are the mechanisms of efficient energy dissipation of the different wave mode types, present in magnetic structures. 
From the theoretical point of view, wave mode transformation theory has been frequently invoked (Cally 2006; Schunker and Cally 2006). According to this theory, acoustic waves (or $p$-modes) propagating from sub-photospheric regions suffer conversion into fast and slow magneto-acoustic waves at the equipartition layer.
This way, a part of the low-frequency $p$-mode energy can escape to the chromosphere in the form of the slow magneto-acoustic mode along the inclined magnetic field lines (Jefferies et al. 2006; Stangalini et al. 2011), frequently called ``magnetic portals''. Since these slow magneto-acoustic waves propagate field aligned in the $\beta <1$ atmosphere,  they ``see'' the local cutoff frequency reduced through the ramp effect as the gravity is lowered by $g\rightarrow g\cos(\theta)$. The fast magneto-acoustic mode in the $\beta <1$ atmosphere will refract due to the gradients of the Alfv\'en speed, and its energy will return to the sub-surface layers. The efficiency of the mode transformation mechanism in transmitting the wave energy to the chromosphere depends on the relative inclination between the wave propagation direction and the direction of the magnetic field, being most efficient for the fields inclined by about 30 degrees (Cally 2006; Schunker and Cally 2006). 
As far as we concerned about the tube waves (see Section 3 also), in the regions of the magnetic portals with low beta plasma, the effective gravity on a particular magnetic field will be modified by the cosine of the angle  subtended by that field line with respect to the direction of the gravity, further causing reduced cut-off and allowing the wave propagation (McIntosh and Jefferies 2006). More specifically, the effects of inclination, magnetic field, and non-uniformity of the medium  along with the gravity stratification lead reduction of the cut-off and propagation of the waves (Spruit and Roberts 1983; Afanasyev and Nakariakov 2015). 


Alfv\'en waves are one of the best candidates to bring the energy to the upper atmosphere because they do not shock at lower layers, and due to their incompressibility they are not affected by damping through viscosity or radiation (see Section 3). Efficient damping of these waves can be achieved through ion-neutral effects in the partially ionized chromospheric plasma (e.g. ambipolar diffusion mechanism,  as discussed by Goodman (1996, 2011); Goodman and Kazeminezhad 2010; Song and Vasyli{\={u}}nas (2011, 2014); Shelyag et al. 2016; Mart{\'\i}nez-Sykora et al. 2016, see below). One of the ways of producing Alfv\'en waves is through the geometrical mode transformation. At heights where fast magnetic waves refract and reflect (typically located in the upper photosphere and chromosphere), these can be partially converted into Alfv\'en waves (Cally and Goossens, 2008; Cally and Hansen, 2011). The most efficient  conversion is achieved for strongly inclined magnetic fields as those in sunspot penumbra or the upper chromosphere at the interiors of the network elements. Such a typical scenario of the Alfv\'en wave propagation in the large-scale chromosphere above a sunspot was recently observed by Grant et al. (2018).


The wave energy transport is extensively studied in observations. A number of studies points that the mode transformation mechanism is indeed acting in the Sun (Moretti et al. 2007; Rajaguru et al. 2013, 2019; Kontogiannis et al. 2014, 2016; Grant et al. 2018). In particular, one of the most prominent phenomena that is now believed to be essentially due to the mode transformation is the presence of high frequency acoustic halos surrounding active regions (Khomenko and Collados 2009; Rijs et al. 2016). According to Moretti et al. (2007),  the same type of magneto-acoustic wave is required to explain both the phenomena of $p$-mode absorption in sunspots and power halos at magnetic canopies. Kontogiannis et al. (2014, 2016) showed that the measured magneto-acoustic wave power in the quiet Sun depends on the magnetic field inclination as predicted by theoretical models of the mode conversion. Long-period waves were observed to be channeled to the chromospheric layers along the inclined magnetic field lines, while the short-period waves were refracted and reflected back at the inclined canopy of a network region, producing magnetic shadows. This behavior certainly affects the amount of the wave power reaching the chromosphere. Rajaguru et al. (2019) demonstrated that the energy flux of acoustic waves within magnetic structures, both in active and quiet regions, peaks around magnetic inclination of 60 degrees, in agreement with the action of the mode-conversion process. 
Mode conversion was also observed to be important for driving Type I spicule oscillations (Jess et al. 2012). Their data provide evidence for magneto-acoustic oscillations, propagating from the surface to above undergoing longitudinal-to-transverse mode conversion into waves at twice the initial driving frequency. The  energy flux to the chromosphere was estimated to be $3\times 10^5$ W m$^{-2}$ which is sufficiently high even to heat the corona and to accelerate the solar wind. Some observational inference of such a mode conversion above an EUV bright point is reported by Srivastava and  Dwivedi (2010).
Direct observational confirmations of the conversion to Alfv\'en waves are still missing. Nevertheless recently, Grant et al. (2018) used HMI/SDO data to provide observational evidences of Alfv\'en waves heating chromospheric plasma in a sunspot umbra.  The velocity showed tangential signatures, and the observed temperature enhancements were suggested to be consistent with mode-converted Alfv\'en waves, after their conversion from magneto-acoustic waves.



Many observational works have been dedicated to try to detect  heating process related to traveling waves of different nature. Carlsson and Stein (1997) demonstrated that the transient bright grains seen in the core of Ca II H and K lines are due to acoustic waves forming shocks. Beck et al.(2009) found that the acoustic flux in the quiet areas can maintain the temperature of semi-empirical models only below 500 km, but is insufficient at heights above $800$-$1200$ km. The  temperature in the quiet chromosphere oscillates between an atmosphere in radiative equilibrium and one with a moderate chromospheric temperature rise, and horizontal canopy structure reflects itself in temperature maps at heights in the low chromosphere (Beck et al. 2013). Using Ca II 853.2 IBIS data, Abbasvand et al.(2020a) found that the deposited magneto-acoustic wave energy balances 30-50\% radiative losses in the quiet chromosphere and 50-60\% of the losses in a plage with vertical field, rising up to 70-90\% in the plage regions with inclined field. This way, significant portions of the radiative losses could be compensated via acoustic wave flux which has been previously converted to magneto-acoustic wave flux in the regions with the inclined fields. Abbasvand et al.(2020b) concluded that the flux contained in magneto-acoustic waves with frequencies up to 20 mHz is sufficient to balance the radiative losses of the quiet chromosphere up to 1000-1400 km height. The difference with the previous results can be attributed to the existence of magnetic shadows (areas with reduced wave power at a given frequency), which prevents part of the acoustic energy reaching chromospheric heights in the network and plage regions, similar to observations by Kontogiannis et al. (2014, 2016). It is worth noting that the quiet-Sun magnetic network elements are surrounded by such a $"$magnetic shadows$"$ (McIntosh and Judge 2001), which are the regions that lack oscillatory power at higher frequencies. The magnetic shadows most likely correspond to the typical $"$acoustic halos$"$, which are the locations of the increased high-frequency ($>$3.3 mHz) power in the solar photosphere (Muglach et al. 2005). Rajaguru et al. (2019) shown that low-frequency waves in the range of 2-4 mHz can still channel a significant amount of energy to the low chromosphere at locations with relatively vertical magnetic field (their frequencies are below the magnetic-field reduced acoustic cutoff). Propagation of these waves is probably assisted by the radiative transfer-related mechanisms as pointed out by Khomenko et al.(2008b). These waves transport up to 2.6 kW m$^{-2}$ to the chromosphere, which is about twice larger than estimated previously by Jefferies et al. (2006). Straus et al. (2008), using high-cadence IBIS measurements at mid-photospheric heights found up to 5 kW m$^{-2}$ energy flux in low-frequency gravity waves.
%


There is still no agreement if high-frequency (magneto-) acoustic waves can supply enough energy to maintain a chromosphere (Fossum and Carlsson 2005; Cuntz et al. 2007). Theoretical calculations by Musielak et al. (1994) reveal that acoustic flux has a broad maximum in the high frequency region around 100 mHz (periods $\approx$ 10 s). Ulmschneider et al. (2005) re-considered  chromospheric heating by short-period acoustic waves arguing that one-dimensional simulations are inadequate since shock merging destroys much of such short period waves. Firm detection of the high-frequency part of the solar oscillation spectrum is still uncertain. According to Cuntz et al. (2007) the high-frequency energy flux was underestimated by Fossum and Carlsson (2005) due to the limited sensitivity of the TRACE data and for not fully assessing the three-dimensional chromospheric magnetic field topology. Using high resolution data by GFPI/VTT,
%
%
Bello Gonz{\'a}lez et al. (2009) revealed the presence of significant acoustic flux into the quiet chromosphere of about 3000 W m$^{-2}$ carried by acoustic waves with frequencies in the 5-10 mHz range (about 2/3 of the flux) and in the 10-20 mHz range (about 1/3 of the flux) predominantly at locations above intergranular lanes. These fluxes can contribute to the basal heating of the quiet chromosphere. Almost twice large fluxes were found from the highest resolution IMaX/SUNRISE data by Bello Gonz{\'a}lez et al. (2010). These results are in contradiction to the earlier work by Fossum and Carlsson (2006) and Carlsson et al. (2007) using TRACE and Hinode data, respectively, but are in agreement with Cuntz et al. (2007).

When considering high-frequency waves and shocks, damping or dissipation effects become important. As discussed above, small-scale disturbances can be easily damped in the chromosphere through various non-ideal mechanisms such as viscosity, radiation, or electrical resistivity. Additionally, in the partially ionized chromosphere ion-neutral effects effectively assist the wave damping. In this regard, numerous works have been aimed at modeling strongly non-linear wave dynamics in the chromosphere in the presence of neutrals. At temporal and spatial scales larger than ion-neutral collisional scales the effect of neutrals can be taken into account using a single-fluid MHD like approximation. In this approximation, the ion-neutral interaction is expressed via the ambipolar diffusion mechanism (Spitzer 1962). The ambipolar diffusion has been studied extensively in the context of chromospheric heating and structure formation, both using the analytical theory or idealized, and even realistic, numerical simulations (see the review by Ballester et al. 2018). It has been shown that ambipolar diffusion allows to dissipate into heat incompressible magnetic waves (e.g., Alfv\'en waves; Khomenko et al. 2018; Gonz{\'a}lez-Morales et al. 2020). Resistive dissipation of Alfv\'en waves, enhanced through ion-neutral interaction, has been considered by Song \& Vasyli{\={u}}nas(2011); Tu and
Song (2013); Shelyag et al. (2016); and Mart{\'\i}nez-Sykora et al. (2016). Arber et al. (2016) argued that viscous damping of shock waves has more importance than ambipolar dissipation.

In the upper chromosphere, collisions may not be strong enough and the difference in the velocities of ions and neutrals (or between the different types of ions or neutrals themselves) may reach some fraction of the sound speed. In this situation, a simplified single-fluid treatment may not be sufficient and multi-fluid modeling should be applied. In a multi-fluid approach, the heating due to neutrals is expressed via the frictional heating term (Braginskii 1965; Leake et al. 2014). The influence of multi-fluid effects on the formation and dissipation of chromospheric shock waves has been scarcely studied, see e.g., Hillier et al. (2016); Snow and Hillier (2019), Snow and Hillier (2020).  Hillier et al. (2016) modeled the formation and evolution of slow-mode shocks driven by reconnection in a partially ionized plasma. A complex multi-fluid structure of the shock transition was modeled revealing structures similar to $C$-shocks or $J$-shocks in the classification by Draine and McKee (1993). The frictional heating associated with ion-neutral decoupling at the shock front was up to 2\% of the available magnetic energy.
In a recent study, Popescu Braileanu et al.(2019a,b) have shown how the ion-neutral decoupling visibly affects waves in neutrals and charges in the chromosphere, and how these effects become more pronounced in shocks. Realistic multi-fluid modeling still remains a challenge due to its complexity, see recent works by Maneva et al. (2017); Ku{\'z}ma et al. (2019). 
Apart from typical chromospheric shocks, recently Srivastava et al. (2018) have observed the presence of pseudo-shocks around a sunspot. Unlike a typical shock as described above, pseudo-shocks exhibit discontinuities only in the mass density. A two-fluid numerical simulation reproduces such confined pseudo-shocks with rarefied plasma regions lagging behind them. It was conjectured that these pseudo-shocks carry an energy of 10$^{3}$ W m$^{-2}$, which is enough to locally power the inner corona and also generate bulk mass flows ( 10$^{-5}$  kg m$^{-2}$ s$^{-1}$), contributing to the localized mass transport. If they are ubiquitous, such energized and bulky pseudo-shocks above active regions could provide an important contribution to the heating and mass transport in the overlying solar corona.

In summary, wave heating of the solar chromosphere is a promising mechanism which, despite being proposed long time ago, still provides grounds for discussions. While the chromosphere seems not to be hot at all times, but rather in an intermittent manner, acoustic waves might provide a basal heating in the quiet areas. Hypothesis of magneto-acoustic and Alfv\'en wave heating are getting more observational evidence as better observations become available. The role of magnetic topology is beginning to be recognized. A particular emphasis for the coming years should be put into the study of the high-frequency part of the spectrum, which also provides new dissipation mechanisms due to partial ionization effects.


\subsection{Energy budget and heating capacity of MHD waves in localized chromospheric flux tubes}\label{Sec:heating_loc_tube}

While, the role of the MHD waves in large-scale solar chromosphere is still illusive and mainly predominant heating by acoustic waves/magneto-acoustic waves is known (see sub-section 4.1),
the recent high-reolution observations detect a variety of MHD modes (kink, sausage, and torsional modes) in localized photospheric and chromospheric tubes (see Section 3), which carry 
substantial amount of energy flux that may balance the chromospheric huge radiative losses, as well as can enable the nascent solar wind plasma flows.
Recent observational findings from high-resolution ground (e.g., Swedish 1-m Solar Telescope (SST), Rapid Oscillations in the Solar Atmosphere (ROSA)) and space-
based (e.g., Interface Region Imaging Spectrograph (IRIS), Solar Dynamics Observatory (SDO)) observations determine the presence of a high amount of energy flux
associated with different MHD modes in different kinds of the magnetic structures coupling the various layers of the solar atmosphere. In the photosphere, wave behaviour ubiquitously detected in pores, sunspots, EUV bight points, tubes related to the plage regions. 
Jess et al. (2009) have reported the detection of Alfv\'en waves with periods of the order of 126-400 s above bright-points with typical energy flux of 1.5$\times$10$^{4}$ W m$^{-2}$ with a 42\% transmission coefficient. Grant et al. (2015) have observed upwardly propagating slow sausage waves in magnetic pores carrying an energy flux of 3.5$\times$10$^{4}$ W m$^{-2}$. These observations infer that magnetic pores transport wave energy to the higher overlying atmosphere, and are also releasing substantial energy in the localized chromospheric plasma for its heating.
The higher order non-axisymmetric modes magnetoacoustic oscillations (m$\geq$1) in the frame-work of both body and surface waves are also observationally detected in magnetic tubes rooted in the solar photosphere (Jess et al. 2017; Stangalini et al. 2018; Keys et al. 2018).

As far as the transverse waves are concerned, Morton et al. (2012) have observed the simultaneous presence of fast kink and sausage waves in mottles and fibrils using
ROSA H$\alpha$ high-resolution observations. They have estimated that these wave modes carry an average energy of respectively 4300 W m$^{-2}$, and 11700 W m$^{-2}$ in the solar chromosphere,
which is still the large amount present in the lower solar atmosphere. Srivastava et al. (2017) have reported the ubiquitous presence of high frequency
($\approx$12-42 mHz) torsional motions in spicular-type structures in the chromosphere. Their numerical model showed that these observations resemble
torsional Alfv\'en waves associated with high frequency drivers containing a huge amount of energy (10$^{5}$ W m$^{-2}$) in the chromosphere. A significant amount of wave energy (10$^{3}$ W m$^{-2}$) is being transferred into the overlying corona from the chromosphere to balance its radiative losses, even after partial reflection of the waves from the TR.
Grant et al. (2018) have reported the evidence of Alfv\'en wave heating of the chromospheric plasma in an active region sunspot umbra in the frame-work of mode conversion and the formation of magnetoacoustic shocks. 
In conclusion, the various drivers in the chromosphere help in generating different
MHD modes in magnetic flux tubes along with a sufficient amount of energy flux. In the chromosphere, such magnetic structures exhibit a significant role in guiding MHD waves, and thus the related 
energy fluxes. However, the filling factor of these localized MHD waveguides in the lower solar atmosphere, and the global evolution of the particular wave mode is not possible eventually posing a question on over-all contributions of these waves in the global heating of the solar corona. However, definitely the waves could play an important role in heating the localized chromosphere and corona.

In the next sub-section 4.3, we will illustrate the heating aspects of the chromospheric prominence plasma.



\subsection{Waves Heating in Prominences}\label{Sec:heating_prom}


Solar prominences are structures with properties which resemble the chromosphere and are embedded in the solar corona. Due to their peculiar physical properties, low temperature and high density compared to that of the solar corona, and dynamic behavior, solar prominences are the subject of intense research. Due to their low temperature, the prominence plasma is partially ionized, although the exact degree of ionization is still unknown and, probably, is not a constant quantity inside each considered prominence. The origin of the prominence mass has been an open question for many years and, nowadays, we know that it must originate in the chromosphere (Pikelner 1971; Mackay et al. 2010), because there is not enough plasma in the corona for their formation (Song et al. 2017). Among the different models that have been proposed to explain how chromospheric plasma becomes prominence material the evaporation-condensation model (Antiochos et al. 1999; Mackay et al. 2010; Luna et al. 2012; Xia et al. 2014)  is the one most rigorously studied for prominence formation. However, all the different models have been developed assuming fully ionized conditions, while partial ionization brings the presence of neutrals and electrons in addition to ions, thus collisions between the different species are possible and the effects on the prominence mass formation must be considered (Karpen et al. 2015). Another key problem in the physics of prominences is how these cool and dense structures are supported in the hotter, more rarefied solar corona. Considering the prominence as a partially ionized plasma, the problem of how neutrals are supported in prominences is a matter of great interest. Bakhareva et al. (1992) pointed out that, after perturbing a 2D Kippenhahn-Schl\"uter magnetic configuration, the system underwent amplified oscillations of density, magnetic field, and velocity which, finally, destroy the prominence. The instability is caused by the
inability of the magnetic field to support the plasma's neutral component against gravity. However, Terradas et al. (2015) solved the two-fluid equations, including ion-neutral and charge-exchange collisions, to study the temporal behavior of a prominence plasma embedded in the solar corona. In their model, the prominence is represented by a large density enhancement, composed of neutrals and ions, and the magnetic configuration is quadrupolar with dips. The results showed that partially ionized prominence plasma can be efficiently supported when the coupling between ions and neutrals is very strong (see also Gilbert et al. 2002, 2007). Summarizing, since prominences are made of partially ionized plasmas, it is very important that we take into account ion-neutral coupling when describing the physics of prominence formation, support and stability.


The heating of solar prominence plasma has been a matter of intense debate. In order to exist, prominences need mechanical equilibrium as well as a detailed energy balance between heating and radiative cooling.  Energy balance studies suggest that incident radiation provides most of the heating of prominence plasma (Gilbert 2015), however, this radiative heating depends on the illumination from the surrounding atmosphere. Moreover, radiative equilibrium prominence models, constructed from a balance between incident radiation and cooling (Anzer and Heinzel 1999; Heinzel et al. 2010; Heinzel and Anzer 2012), as well as differential emission measures have pointed out that a further unknown heating is required in order to reproduce the observed temperatures in the prominence cores (Labrosse et al. 2010; Heinzel 2015; Heinzel and Anzer 2012) and to balance the radiative losses (Parenti and Vial 2007; Parenti 2014). 

On the other hand,  ground- and space-based observations have confirmed the presence of oscillatory motions in prominences, which have been interpreted in terms of standing or propagating MHD waves (Roberts 2000, 2008; Aschwanden 2004; Nakariakov and Verwichte 2005; Nakariakov et al. 2016b; Oliver 1999; Oliver and Ballester 2002; Oliver 2009; Erd{\'e}lyi and Goossens 2011; Ballester 2015; Arregui et al. 2018). Observational studies of small amplitude prominence oscillations have shown the existence of periods which can be distributed in three groups: of a few minutes, between 10--40 min and between 40--90 min. Furthermore, observations have also provided evidence about the damping of prominence/filament oscillations. In most of the observations in which the damping time has been determined, those damping times are between 1 and 4 times the corresponding period, and large regions of prominences/filaments display similar damping times (Arregui et al. 2018).  The observational evidence about wave activity in prominences together with theoretical developments about MHD waves have boosted the development of prominence seismology (Arregui et al. 2018) which allows the determination of the prominence physical parameters from the comparison of observed and theoretical wave properties. However, much work remains to incorporate partial ionization effects, in particular multi-fluid physics, into existing models of oscillatory phenomena in prominences.
 
    Mechanisms based on wave dissipation have been proposed for the heating of the solar atmosphere (see Arregui 2015 for a review), therefore, an additional source of energy to heat prominences could be provided by these mechanisms. Kalkofen (2008) estimated the acoustic flux in the photosphere and found that it is much smaller than the energy radiated by the chromosphere. Therefore, it seems that sound waves are not the source of chromosphere and prominence heating, while Alfv\'en waves could be a good candidate. However, P{\'e}cseli and Engvold (2000) estimated the wave heating associated to  Alfv\'en wave dissipation in prominence plasma finding that the contribution is very small. Parenti and Vial (2007) derived the distribution of nonthermal velocities at various temperatures in a prominence observed with SOHO/SUMER. They assumed that these velocities were the signatures of wave propagation, and computed the energy flux for Alfv\'en and sound waves. Using differential emission measure analysis, they computed the radiative losses in the prominence, and made a comparison between both estimations. They concluded that the radiative losses in the prominence-corona transition region (PCTR) could be compensated by Alfv\'en wave heating. Hillier et al. (2013) made a comparison between power spectra coming from prominence waves with that of photospheric horizontal motions. They found that for frequencies less than 7 mHz, the frequency dependence of the velocity power is consistent with the velocity power spectra generated from observations of the horizontal motions of magnetic elements in the photosphere, suggesting that the prominence transverse waves are driven by photospheric motions. Then, the excited MHD waves at the photospheric level could transport energy from the photosphere to the prominences located in the solar corona.  
 
 Several studies have considered the damping of Alfv\'en waves in partially ionized plasmas, like the solar chromosphere, by means of ion-neutral collisions (Haerendel 1992; De Pontieu et al. 2001; James et al. 2003; Leake et al. 2005; Song and Vasyli{\={u}}nas 2011; Soler et al. 2015a,b), and since prominence plasmas are akin to chromospheric plasmas, wave dissipation in partially ionized plasmas could be the source of the required additional heating. However, it is important to keep in mind that the efficiency of waves to heat the plasma depends on the  efficiency of the dissipation mechanism locally (Soler et al. 2016). In this sense, Khodachenko et al. (2004, 2006) made a qualitative study of the damping of MHD waves in a partially ionized plasma considering viscosity, collisional friction and thermal conduction as potential damping mechanisms.  From this study, they concluded that collisional friction is the dominant damping mechanism for MHD waves in prominences. Using the single-fluid approximation, further studies on the temporal and spatial damping of Alfv\'en and magnetoacoustic waves in unbounded and structured partially ionized plasmas with prominence physical properties, but without computing heating estimations, were made by Forteza et al.(2007, 2008); Soler et al. (2008, 2009, 2010); Carbonell et al. (2010); Barcel{\'o} et al. (2011), while Mart{\'\i}nez-G{\'o}mez et al.(2018) have computed the heating in a three-fluid prominence plasma.
 
With the aim to explore prominence heating by means of damped Alfv\'en waves, first of all, Solar et al. (2016) considered an unbounded homogeneous and partially ionized prominence plasma, with a straight and constant magnetic field, and studied the propagation and spatial damping of linear Alfv\'en waves. The plasma was considered to be composed of partially ionized hydrogen and neutral helium at T $<$ $10^4$ \  K. To study wave propagation, the multifluid formalism (Zaqarashvili et al. 2011c) was used, assuming that ions and electrons formed a single fluid while neutral hydrogen and neutral helium formed two separate neutral fluids. Then, from the linearized equations for Alfv\'en waves, an expression for the evolution of wave energy was derived,
\begin{eqnarray}
\frac{\partial U}{\partial t} + \nabla \cdot \Pi = - Q\, .
\label{evol}
\end{eqnarray}
This equation includes the temporal evolution of wave energy density, {\bf \textit{U}}, corresponding to the sum of kinetic energies of ions, neutral hydrogen, neutral helium and magnetic energy; the divergence of wave energy flux, {\bf \textit{$\Pi$}}, corresponding to the amount of energy propagated by the wave; and the loss of wave energy, {\bf \textit{Q}}, due to dissipation by collisions. As the wave propagates, wave energy is absorbed by the plasma and converted into internal energy, and the quantity, {\bf \textit{Q}}, can be identified as the wave heating rate. The Alfv\'en energy flux can be computed from {\bf \textit{$\Pi$}}, and the time-averaged energy flux for a forward-propagating wave is,
\begin{eqnarray}
\langle \Pi \rangle= \frac{B^2}{2 \mu} \frac{Re(k)}{\omega} V^2 \exp[-2 Im(k) s]\hat e_B\, .
\label{Alflux}
\end{eqnarray}
in $W \cdot m^{-2}$. The time-averaged energy flux is proportional to $\exp[-2 Im(k) s]$, with $Im(k)$ the imaginary
part of the wavenumber and, s, the propagation distance, indicating that the net energy carried by the Alfv\'en wave exponentially decreases, because of collisions, as the wave propagates (see Soler et al. (2016) for details). Then, a characteristic energy absorption length scale, $L_A$, can be defined as 
 \begin{linenomath*}
\begin{equation}
L_A = \frac{1}{2 Im (k)}\, ,
\label{eq:abs}
\end{equation}
 \end{linenomath*}
which represents the length scale at which the energy absorbed is deposited in the plasma.  
Figure 5 (left panel) displays the energy absorption length, $L_A$, computed from Eq.~(\ref{eq:abs}), as a function of the wave period, $P$, for different values of the hydrogen ionization ratio, $\xi_i = {\rho_i}/({\rho_i + \rho_H})$, and typical prominence parameters ($\rho = 5  \cdot 10^{11}$ kg\, m$^{-3}$, 
$T = 8000$ K, $B = 10$ G, and abundance ratio, $A_{He}/A_H = 0.1$). 
As it can be expected, when the hydrogen ionization degree decreases, the wave damping is stronger. Also, Figure 5 (left panel) shows that low period waves are efficiently damped in the plasma, while for long periods waves the damping is less efficient. These results indicate that for waves with periods $P > 100$ \ s, typical of waves in prominences, the wave energy absorption and associated heating are negligible since the energy absorption length is of the order of $10^6$ km, which is unrealistic for prominence's length scales. 
\begin{figure*}
\centering{
\resizebox{14cm}{!}
{\includegraphics[width=5cm, height=5cm]{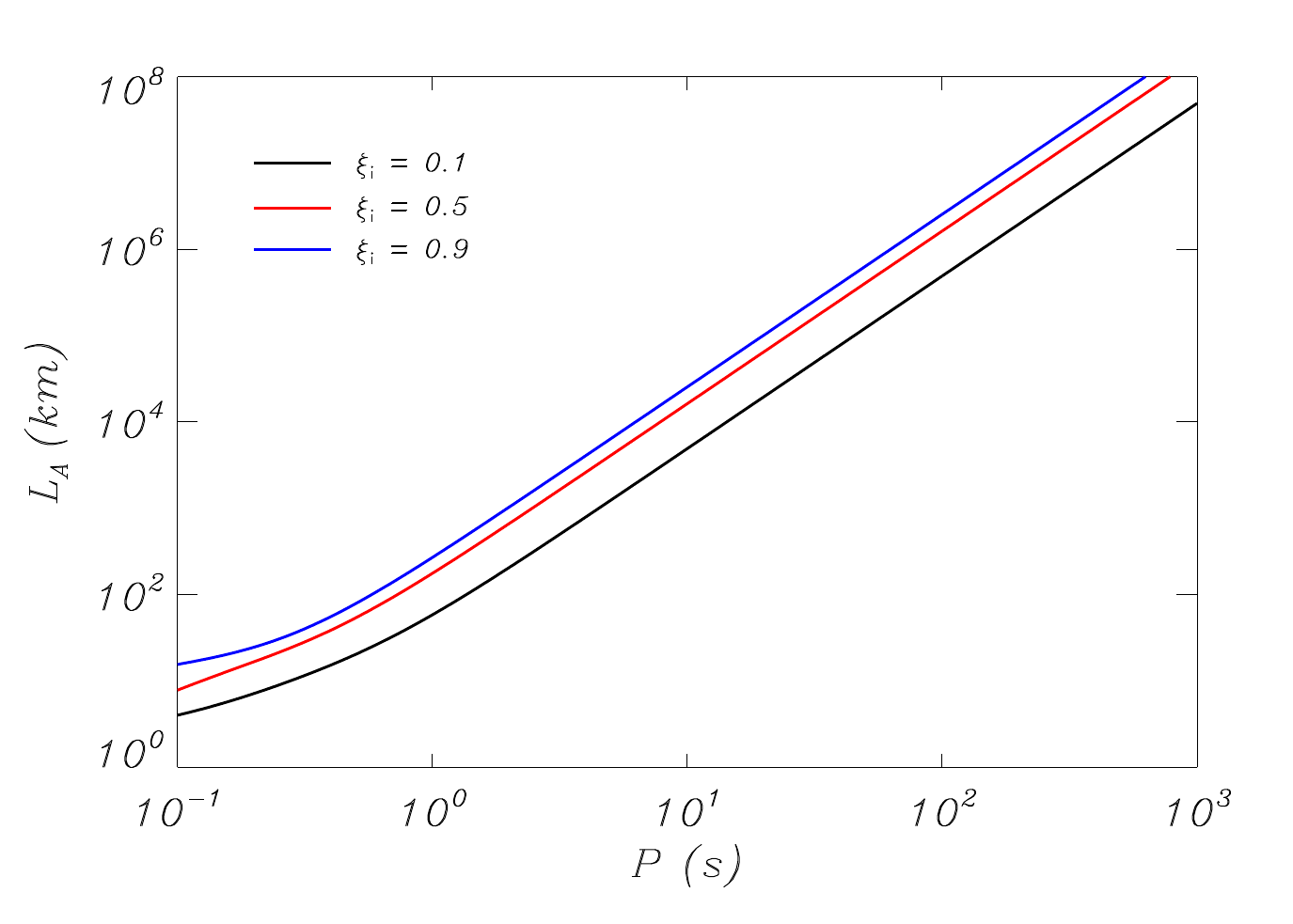}
{\includegraphics[width=5cm,height=5cm]{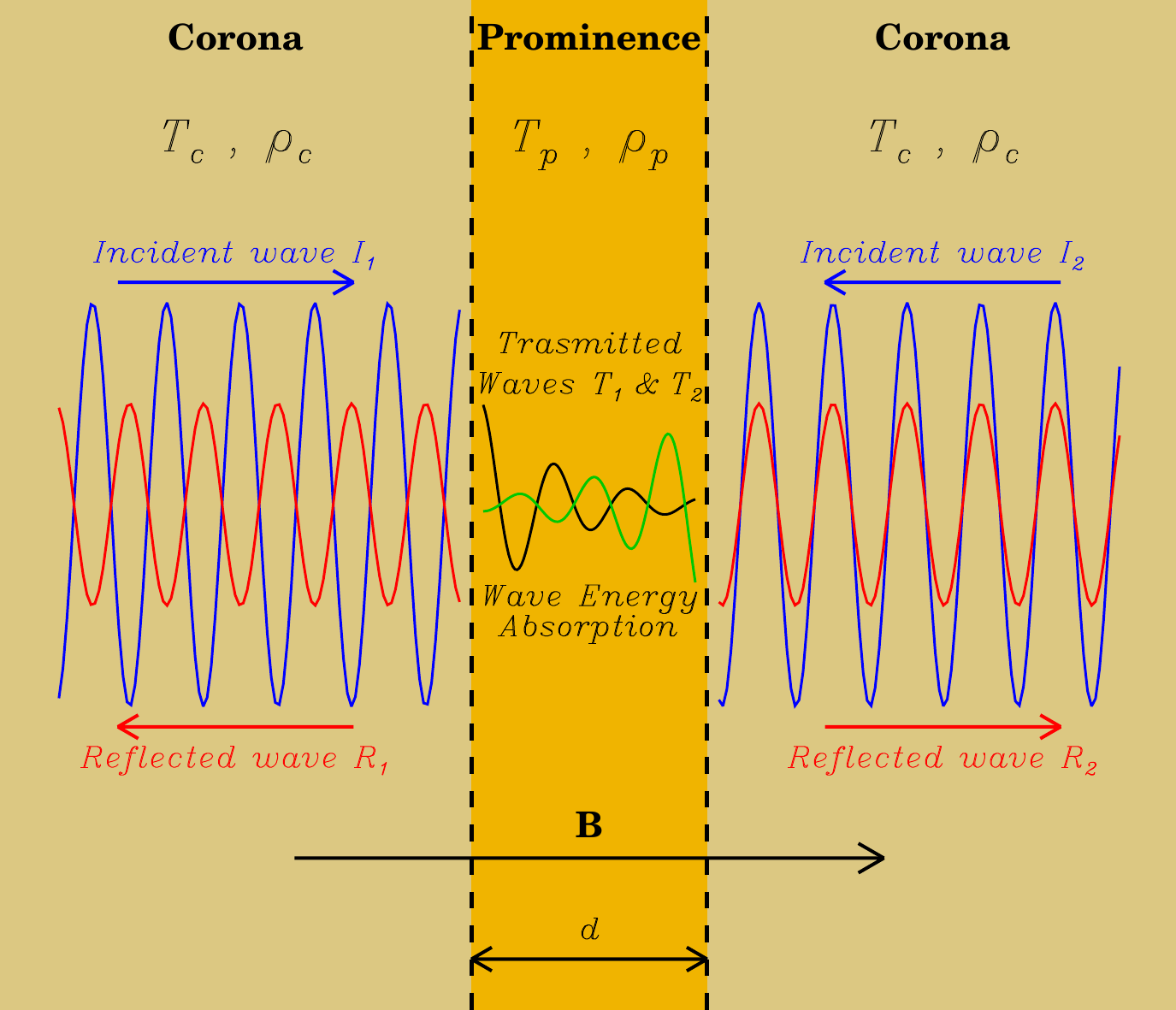}}}}
 \caption{Left panel: Energy absorption length, $L_A$, as a function of the wave period,
$P = {2 \pi}/{\omega}$, for Alfv\'en waves propagating in a prominence plasma
when different values of the hydrogen ionization ratio, $\xi_i$, are assumed.
The meaning of the various line styles is indicated within the figure. Logarithmic scale has been
used in both axes. Right panel: Sketch of the prominence slab model. Credit: Soler et al. 2016, A\&A, 592, A28, reproduced with permission \copyright \ ESO.} 
\label{figH1}
\end{figure*}
Next, Soler et al. (2016) investigated whether the energy of Alfv\'en waves incident on a solar prominence can be deposited in the prominence medium because of this damping. To this end, they represented the prominence by a slab with a transverse magnetic field embedded in the solar corona (see Fig.~\ref{figH1}, right panel) and assume the presence of Alfv\'en waves that are incident on the prominence-corona interface. These Alfv\'en waves come from the footpoints of the magnetic field lines which support the prominence, propagating along these field lines.
In solar prominences, the strong discontinuity in density between the prominence and the corona could lead to Alfv\'en wave reflection and decrease the amount of heat released inside the prominence. However, Hollweg (1984) suggested that the decrease of the Alfv\'en wave velocity inside a prominence slab could act as a resonant cavity for incident Alfv\'en waves. Using this approach, Soler et al. (2016) studied the trapping of energy of Alfv\'en waves incident on a prominence and the associated heating rate. The incident Alfv\'en waves on the prominence-corona interface (see Fig.~\ref{figH1}, right panel) are partly transmitted into the prominence slab and partly reflected back to the corona. The condition for the excitation of cavity resonances
is that the frequency of the incident waves matches a natural frequency (eigenmode) of the prominence, then, the trapping of wave energy within the prominence slab can be very efficient when the period of the incident Alfv\'en waves matches a resonance period. The channeled energy into the prominence is then dissipated by ion-neutral collisions, and can efficiently contribute to the heating of the plasma. The heating rate within the slab caused by the Alfv\'en wave dissipation, {\bf \textit Q}, is computed using the ion velocity amplitude of the waves transmitted into the slab. First of all, this heating rate is time-averaged over one wave period and, next, it is spatially averaged within the slab. The obtained volumetric heating rate depends on the power spectrum of the waves incident on the prominence which, unfortunately, is not well known from observations and needs to be assumed (see below). Finally, the total heating produced by a broadband driver is obtained by integrating the volumetric heating rate over a range of periods. Then, to estimate this total heating rate, Soler et al.(2016) used two different power spectra: a flat power spectrum with a constant velocity amplitude of $10$ km/s, and a velocity amplitude which decreases with a power law  following the observations by Hillier et al. (2013) (see Fig.~\ref{figH2}a and b), and compared the obtained heating rates with the prominence radiative losses. Using different approaches, and for typical prominence parameters, the radiative losses from the prominence slab were estimated to be in the range $10^{-5}$ W\,m$^{-3}$ and $10^{-4}$ W\, m$^{-3}$. The comparison of the integrated heating rates of Fig.~\ref{figH2} with the estimated radiative losses indicates that wave heating may compensate for a non-negligible fraction ($\sim 10\%$) of the energy lost by radiation. Therefore, the estimated wave heating rate represents a small contribution to the total heating necessary to balance the emitted radiation, but it can possibly account for the additional heating necessary to explain the observed prominence core temperatures (Heinzel et al. 2010). 
This model is rather simple, but it is a first attempt to try to understand how the damping of Alfv\'en waves due to dissipation by collisions can contribute to prominence heating. In order to develop more accurate and efficient prominence heating mechanisms, realistic 2D or 3D numerical models of prominence support, prominence fine structure, ionization equilibrium and energy balance are needed as initial configurations for further studies.  

In general, the fluid (magnetized or non-magnetized) systems are subjected to the arbitrary perturbations. There are likely physical scenario that either the fluid system evolves waves and oscillatory phenomenon, or it will experience the instabilities. In the next Section~5, we describe the magnetic instabilities that arise in the structured chromospheric plasma.
\begin{figure*}
\centering{
\resizebox{14cm}{!}
{\includegraphics[width=5cm]{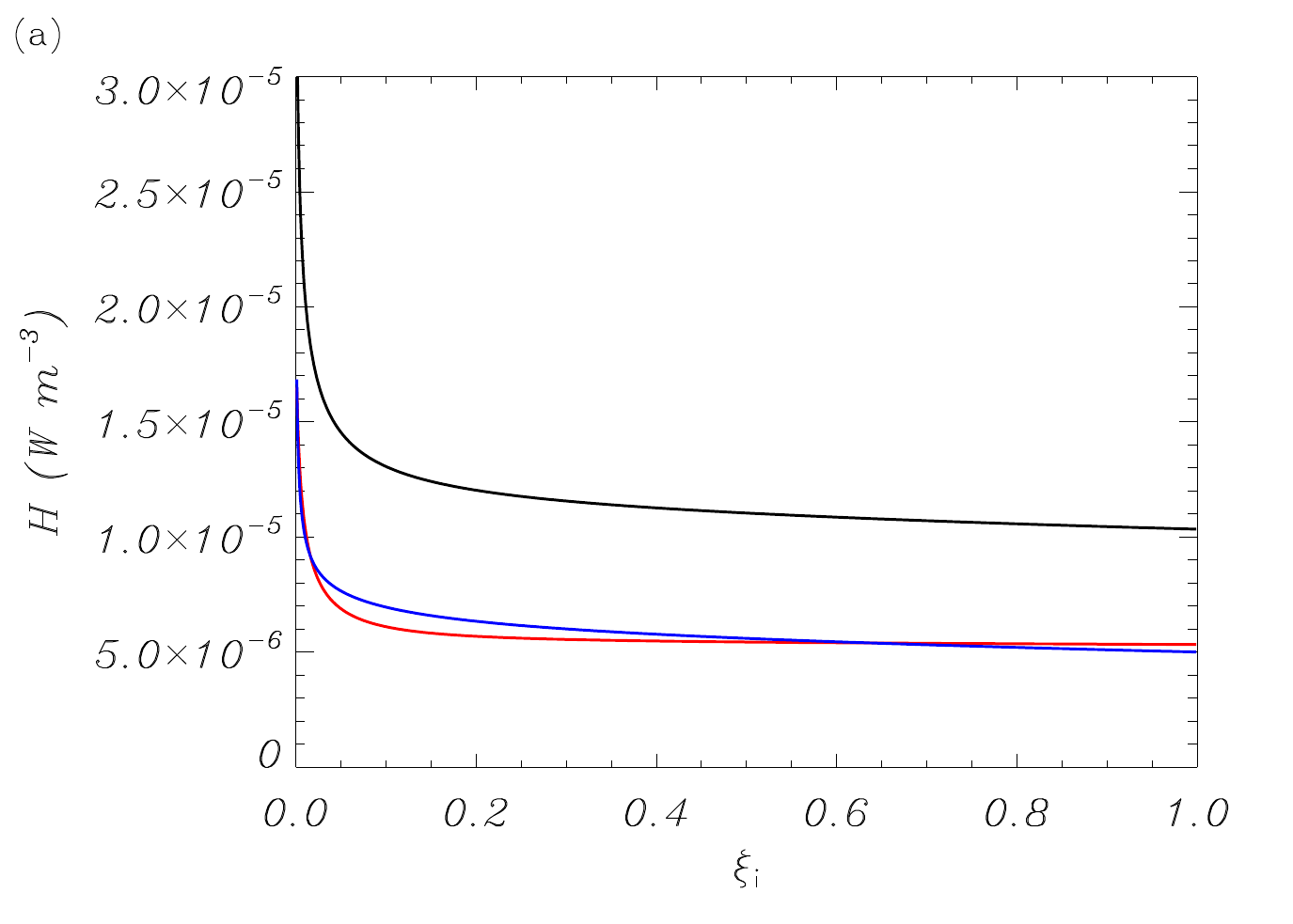}
{\includegraphics[width=5cm]{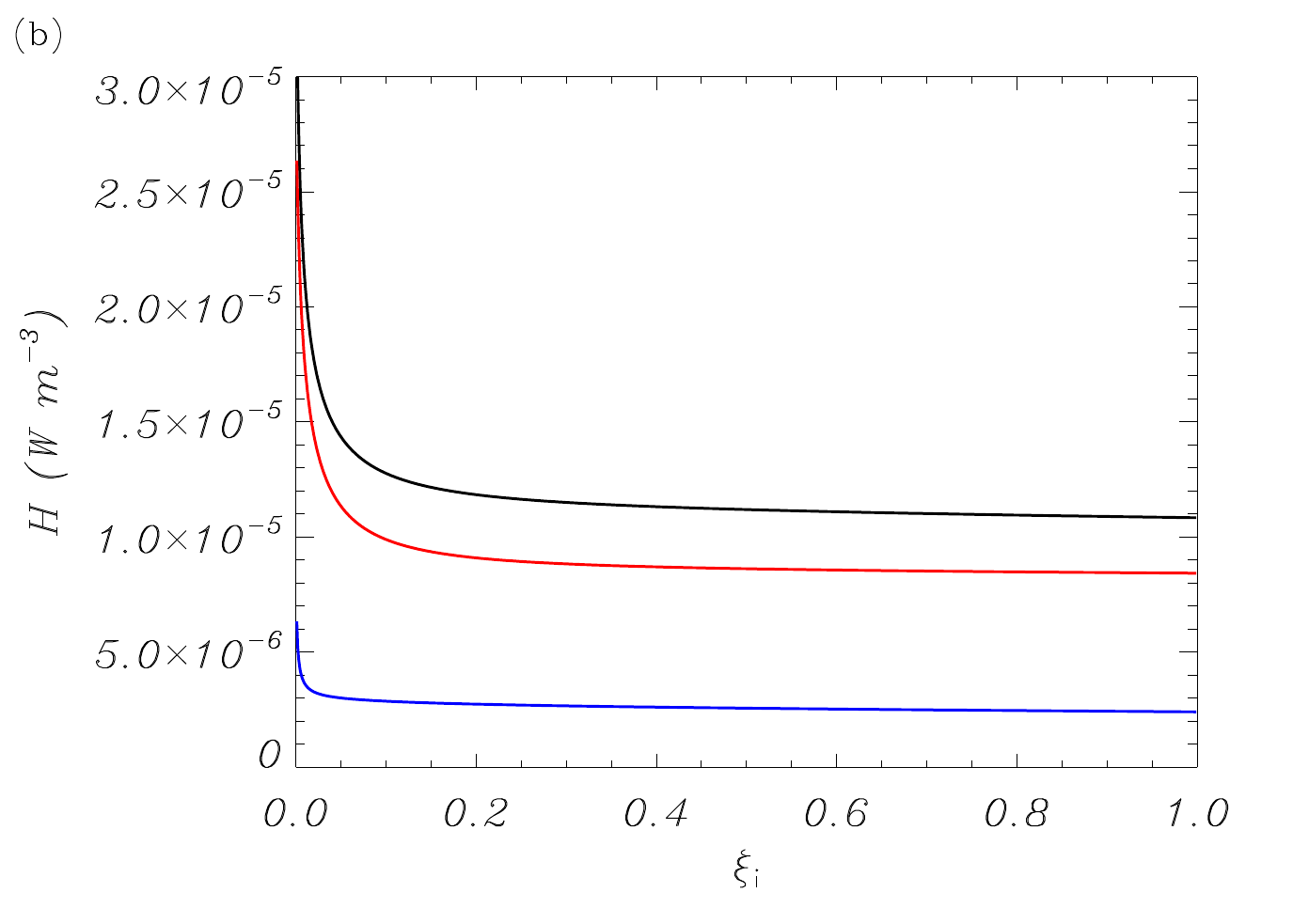}}}}
\caption{a) Total integrated heating, $H$, as a function of the hydrogen ionization ratio, $\xi_i,$ when the velocity amplitude of the incident waves is
$10$ km s$^{-1}$. The black line indicates the full result, the red line indicates the contribution of short periods (0.1 s $\le$ P $\le$ 1 s), and the blue line
indicates the contribution of intermediate periods (1 s $\le$ P $\le$ 100 s). b) Same as panel a) but for the velocity power law of Hillier et al. (2013). Credit: Soler et al. 2016, A\&A, 592, A28, reproduced with permission \copyright \ ESO.}
\label{figH2}
\end{figure*}
%
%
\section{MHD Instabilities in the Chromospheric Plasma}
While wave dynamics is potentially influenced by the structured chromosphere, there may be certain physical conditions that may lead to a subsequent growth of the perturbations in the form of various instabilities. The most common instabilities that are observed and studied in the large-scale chromospheric structures $"$(e.g., solar prominences)$"$ are driven by respectively the gravity and sheared flow: (i) Rayleigh-Taylor (RT) and (ii) Kelvin-Helmholtz (KH) instabilities (Berger et al. 2008; Ryutova et al. 2010; Berger et al. 2010; Hillier et al. 2011; Berger et al. 2017; Mishra and Srivastava 2019). 
The 
RT 
instability was originally proposed by Rayleigh (1899) and Taylor(1950) to 
explain 
the growth of small-amplitude perturbations at 
the interface between two fluids. 
This instability takes place when a denser fluid caps 
a lighter fluid, while the interface between these fluids is perturbed against the gravity. According to Priest (2014), 
the dispersion relation for the uniformally magnetized fluid 
can be expressed as 
\begin{equation}
\omega_{th}^{2}=-gk{\frac{(\rho_{h}-\rho_{l})}{(\rho_{h}+\rho_{l})}}+{\frac{2B^{2}k_{x}^{2}}{\mu(\rho_{h}+\rho_{l})}}\, .
\end{equation}
Here $g$ is the gravitational acceleration of the Sun, $B$ is the horizontal component of the magnetic field, 
$k={2\pi}/{\lambda}$ is the wavenumber with 
$\lambda$ being the characteristic wavelength of the RT instability.
The symbols $\rho_{h}$ and $\rho_{l}$ denote the densities of the heavy and lighter fluids, respectively. 
The dispersion relation as given in Eq.~21 describes that in the absence of the magnetic field the interface will be RT unstable as ${\omega_{th}}^{2}<0$.
providing the condition that heavy fluid ($\rho_{h}$) lies above lighter one ($\rho_{l}$). The oscillations of the interface along the magnetic field
renders the stabilising effects as the second term in Eq.~21 becomes positive.
A representative evolution of RT instability in the solar  prominence system, resulting from the above equation, is shown in Figure 7 both in the observations (left) and numerical modeling (right). 
\begin{figure*}[!t]
\mbox{
\includegraphics[width=8cm]{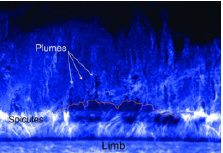}
\includegraphics[width=8cm]{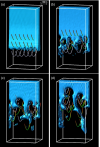}
}
\caption{Left: The 30th November 2006 Ca~II H-line prominence observed on the west limb at 04:50:10 UT, in which 
RT instability-related plumes, upflows, and cavities are well seen (image adapted from Berger et al., 2010). Right: The numerically simulated 
RT instability evolution in form of the growth of plumes/bubbles are seen, which match the observational data (image adapted from Hillier et al. 2012). 
$\copyright$ AAS Reproduced with permission.}
\label{fig:torsional}
\end{figure*}
The evolution of the bubbles and plumes are characteristic features associated with the RT instabilities evolved in the solar chromospheric prominence plasma (Berger et al. 2010; Hillier et al. 2012; Mishra and Srivastava 2019). 

The linear theory of the 
RT 
instability in the partially ionized plasma was developed by D{\'\i}az et al. (2012). 
Additionally, Khomenko et al.(2014a) showed that the non-ideal, partially ionized chromospheric plasma with resistive and ambipolar diffusion taken into account experiences larger growth rate of the RT instability (RTI), faster plasma flows, rapid downflows, evolution of the high temperature bubbles, and the asymmetry between large rising  bubbles and small-scale down-flowing fingers, compared to the plasma evolution in 
the framework of ideal MHD. Such dynamical plasma processes, elevated temperatures, and RT features are well observed by Mishra and Srivastava (2019). Usually, the ideal MHD is sufficient to explain the RT instability. However, in order to explain generation of fine structures, their dynamics, and plasma heating the non-ideal effects are required.

\begin{figure*}[!t]
\centering
\includegraphics[width=10cm]{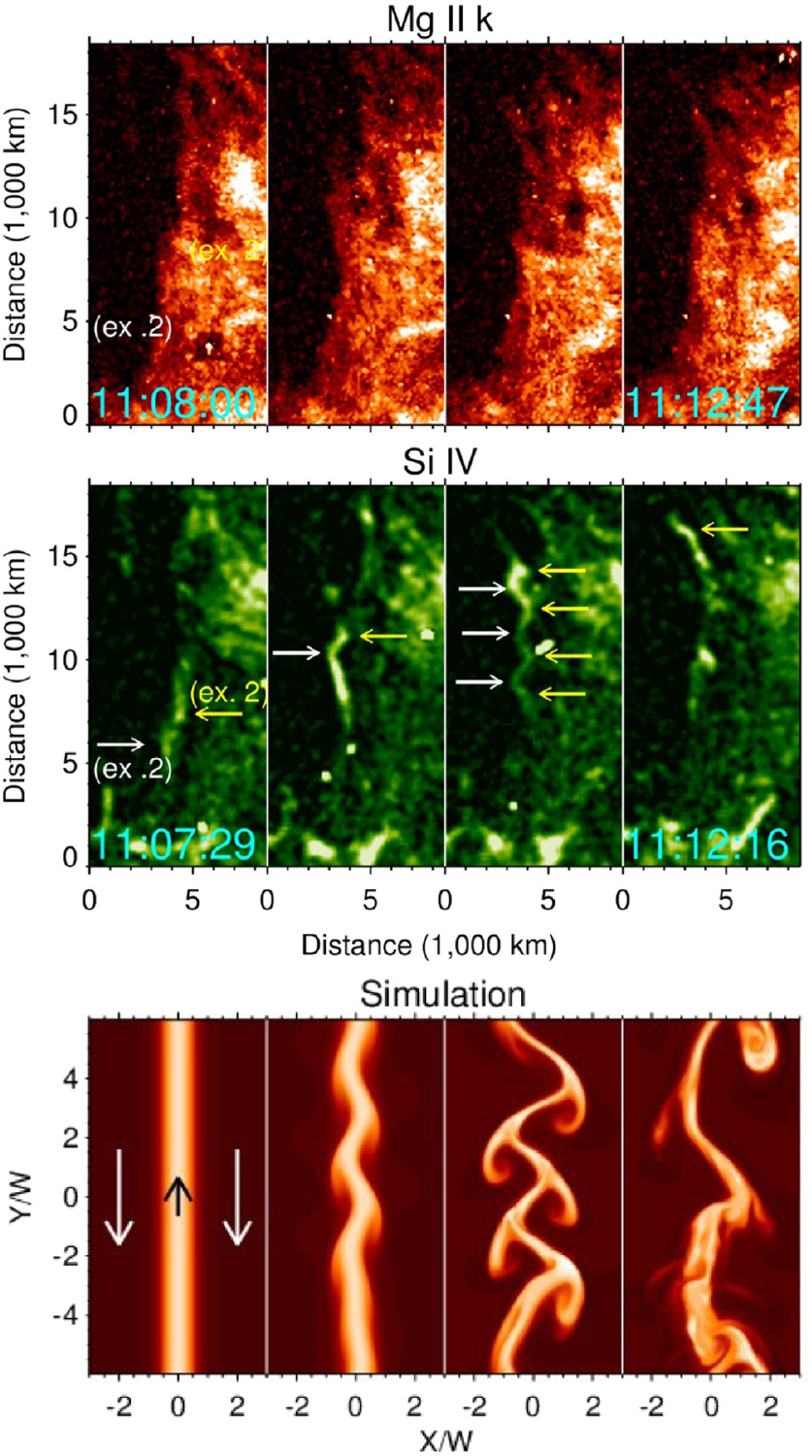}
\caption{
The observational data of the KH unstable upflowing plasma in the solar prominence (top and middle) and 
the results of MHD numerical simulations of the instability driven by shear motion at the interface between two fluids (bottom) (Image adapted from Hillier and Polito 2018).
$\copyright$ AAS \ Reproduced with permission.}
\label{fig:torsional}
\end{figure*}
KH instability takes place 
in a medium with a sufficiently strong sheared flow. 
In a magnetic-free fluid 
the dispersion relation 
for the KH 
instability 
resulting at the interface which separates 
two different states of a gas
is given as Chandrasekhar (1961) and Choudhuri (1998):
%
\begin{equation}
\frac{\omega_{th}}{k}=\frac{\rho_{1}U_{1}+\rho_{2}U_{2}}{\rho_{1}+\rho_{2}}\pm\left[\frac{g}{k}\frac{\rho_{1}-\rho_{2}}{\rho_{1}+\rho_{2}}-\frac{\rho_{1}\rho_{2}(U_{1}-U_{2})^{2}}{(\rho_{1}+\rho_{2})^{2}}\right]^{\frac{1}{2}}
\end{equation}

%
where $U_{1}$ and $U_{2}$ are gas velocities 
at both sides of the interface 
and $\rho_{1}$ and $\rho_{2}$ the corresponding mass densities. 
When, the expression within the square root of Eq.~22 becomes negative then,
\begin{equation}
\rho_{1}\rho_{2}(U_{1}-U_{2})^{2}>({\rho_{1}}^{2}-{\rho_{2}}^{2})\frac{g}{k}
\end{equation}
depicts the evolution of the Kelvin-Helmholtz instability when the two fluid layers
move with the different speeds (Choudhuri 1998). Perturbations with the small 
wavelenghts becomes unstable when the velocity difference in both the layers is small. 
These basic equations (Eqs. 21 and 22) respectively for RT and KHI instabilities can be applied in the incompressible limit.
In the presence of the magnetic fields (B1 and B2), the inset condition for the K-H instability will be (Chandrasekhar 1961; Yuan et al. 2019):
\begin{equation}
\left[\vec{k}.\Delta \vec{U} \right]^{2}=\frac{\rho_{1}+\rho_{2}}{\mu(\rho_{1}\rho_{2})}\left[(\vec{k}.\vec{B_{1}})^{2}+(\vec{k}.\vec{B_{2}})^{2} \right].
\end{equation}
Here, $\vec{k}=2\pi/\lambda$ is the wave vector, $\lambda$ is the characteristic wavelength of the perturbations, and $\Delta \vec{U}$ is the velocity 
difference in fluid layers.

It is noteworthy that the KH instability is the most common instability that is observed and modeled in the cool chromospheric plasma, cool jets, and in spicules (Zhelyazkov et al. 2015; Kuridze et al. 2016; Antolin et al. 2018).
An example of the unstable upflows in the Mg II k and Si IV channels is observed by Interface Region Spectrograph (IRIS) in the solar prominence (Figure 8, top and middle panels), while the simulation of the resulting KH instability is displayed in the bottom panel of Figure 8, that arises due to the counter-propagating flows generating the shear motion at the fluid interface (Hillier and  Polito 2018). In the example shown in Fig.~8, in the middle, we note that the instability leads to the growth of perturbations with a specific parallel wavelength. Moreover, simulations illustrated in the bottom panels of Fig.~8, display the appearance of coherent structures, rather than the nonlinear cascade to shorter wavelengths. The one most important implications of KH instability may be that it can result in turbulent flows in the chromospheric plasma, however, it is theorized in the observations of thin and cool chromospheric jets (Kuridze et al. 2016). 
The KHI was studied in the partially ionized plasma by Soler et al. (2012). The analysis of the instability in the partially ionized plasma shows that the ion-neutral collisions may lead to the fast heating of the generated KH instability vortices (Soler et al. 2012). 
These vortices, in the presence of non-ideal effects, may have a capability to heat the chromospheric plasma. Using simple analytical methods, Ballai et al. (2019) have shown that the dissipative instabilities appear for the flow speeds that are lower than the KHI threshold at the fluid interfaces. While viscosity tends to destabilise the plasma, the effect of partial ionization through the Cowling resistivity will act towards stabilising the interface. However, more observations and 
numerical simulations have to be performed to improve our knowledge 
in this area. 
%
\begin{figure*}[!t]
\centering
\includegraphics[width=12cm]{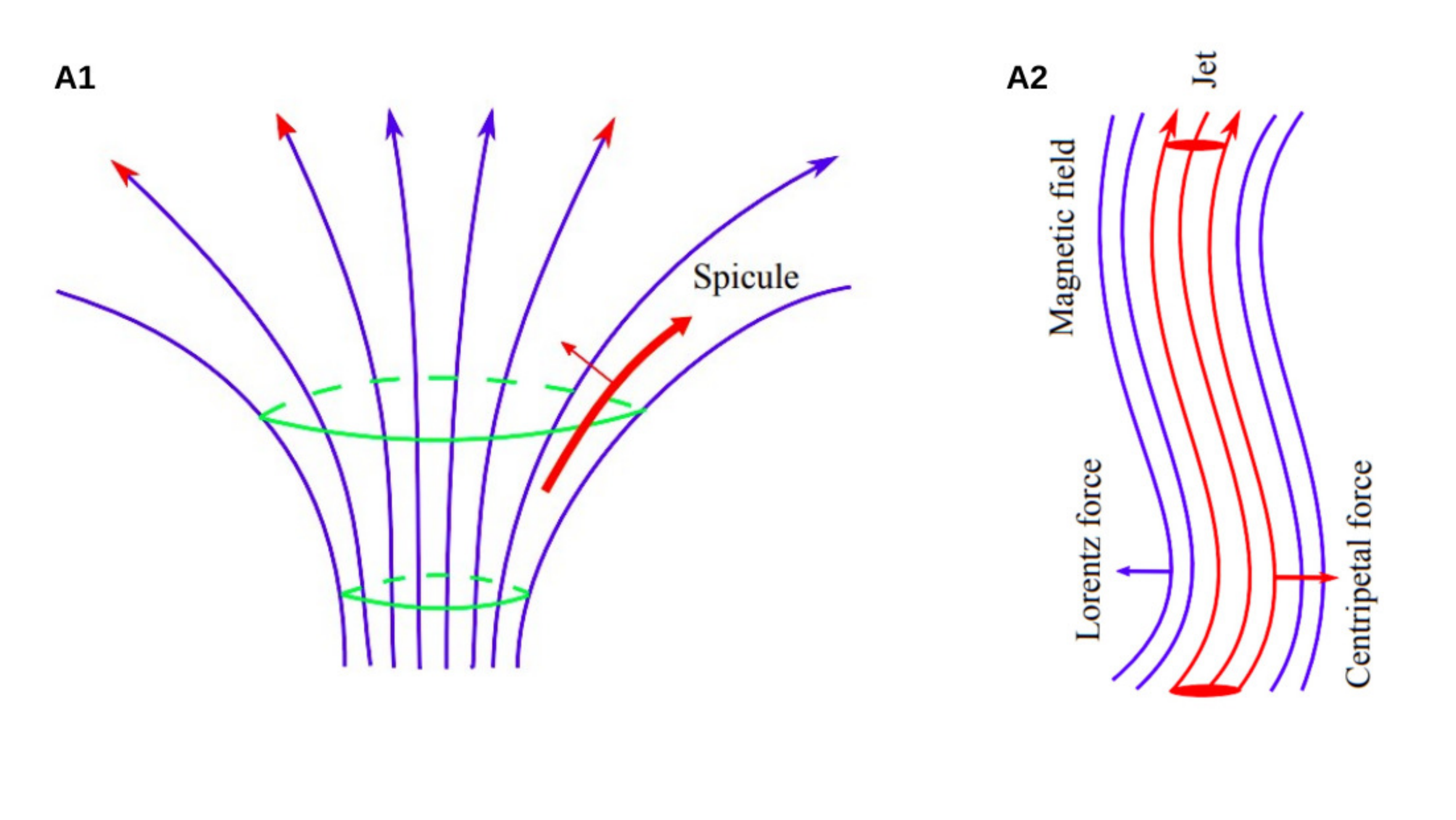}
\includegraphics[width=10cm]{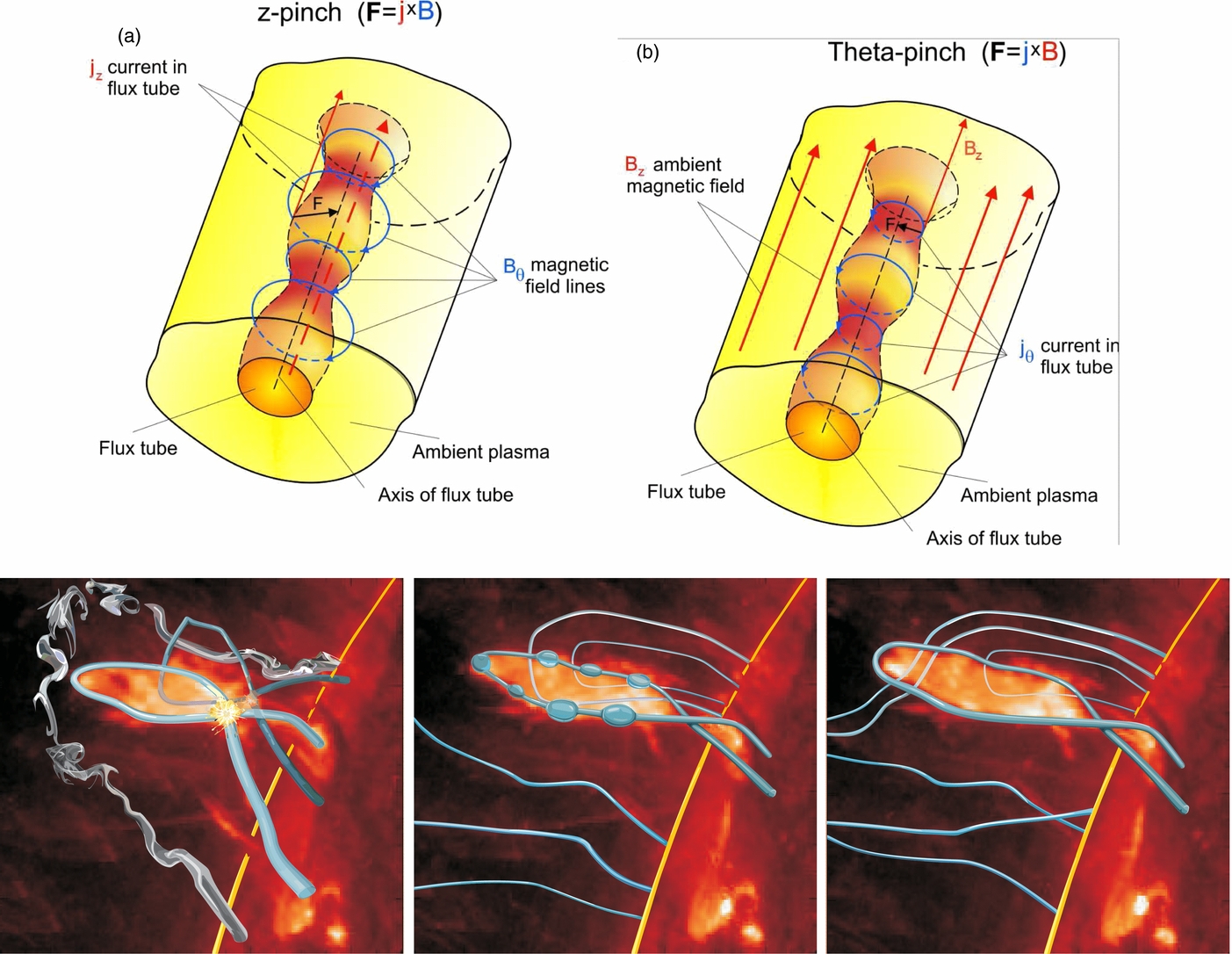}
\caption{Top: The schematic view 
of the evolution of 
kink
instability in the chromospheric jets (adapted from Zaqarashvili et al. 2020). 
Middle: A schematic view related to the physical mechanism of sausage instability in $z$-pinch (panel 'a') and $\theta$-pinch (panel 'b'), 
shown in the context of Sun's magnetized plasma. 
Bottom: 
The sketch of the sausage instability (middle-panel) 
in an active prominence on 12th September 2011 
overlaid on SDO/AIA 304 \AA~image (image adapted from Srivastava et al. 2013). $\copyright$ AAS \ Reproduced with permission. }
\label{fig:torsional}
\end{figure*}

Cylindrical jets are subject to the $m=1$ kink instability which  depends on the Alfv\'en Mach number (specified as the ratio of Alfv\'en and flow speeds) (Zaqarashvili et al. 2020). The kink instability is related to the flow instability (i.e. KHI, see above), where the centripetal force increases the transverse displacement of the jet when it overcomes the Lorentz force of magnetic field lines (panel A2 of Figure 9). It is shown that the 
kink instability may by important for spicules that rise up at the peripheries of vertically expanding magnetic flux tubes (panel A1 of Figure 9) (Zaqarashvili et al. 2020). Therefore, inclined spicules may be more unstable and have higher transverse speeds. Periods and growth times of  unstable modes are comparable to the life time of type II spicules.  Therefore, it may indicate an interconnection between high-speed flows and the rapid disappearance of type II spicules in chromospheric spectral lines (Sekse et al. 2012). 

Chromospheric magnetic flux tubes may also become unstable owing to their twist. Twisted flux tubes can be subject to magnetic kink instability. It is shown that the motion of twisted tubes reduces classical Kruskal-Shafranov instability threshold for the kink instability (Zaqarashvili et al. 2010b). Hence, this instability can be important in spicules. Another possibility is the $m=0$ sausage or pinch instability.  
However, there is no evidence of the sausage instability in the solar atmosphere. 
The only exception can be the one originally excited in the eruptive solar prominence (Figure 9, bottom panel), and also visible in its activation and heating phase (Srivastava et al. 2013). If the azimuthal component of the magnetic field ($B_{\theta}$) at any point in the magnetic flux tube becomes larger than $B_{\theta}$ outside the tube, the enhanced magnetic pressure
makes that point on the tube pinched, and triggers the sausage instability. In the case of the sausage instability related perturbations, the axial field is compressed at the locations inside the flux tube, where pinching generated and the enhanced magnetic pressure there, further opposes the growth of pinching and thus the growth of the sausage instability. In other words, the condition of $B_{z} > B_{\theta}$ 
should be the most general criterion for the stability against the sausage  instability.

Apart from above mentioned standard instabilities, there are several other plasma instabilities that are evolved under specific physical conditions,  and they become highly important in determining the energy and mass transport in the chromosphere and its coupling with other layers of the solar atmosphere. The chromospheric heating may also be linked with the convective overshooting motions in the lower chromosphere driving the "Farley-Buneman (FB) instability" (Fontenla et al. 2008; Gogoberidze et al. 2009; Madsen et al. 2014). Some studies also advocate the liberation of the energy due to the fast magnetic reconnection which develops as a consequence of the "plasmoid instability" without anomalous resistivity enhancements in the middle of the solar chromosphere (Ni et al. 2015). 
Under specific conditions of local thermal equilibrium in the solar corona, the "thermal instability" can also be evolved, which further leads to the initiation of the condensations locally in 
the corona with the plasma material cooling down to the chromospheric  temperature there. Such plasma condensations may form a cool prominence, or evacuate downwards under the influence of the gravity and
gas pressure constituting the coronal rain (Antolin 2020).
%
\section{Discussion, Conclusions, and Some Outstanding Questions}
The complexity of the plasma conditions in the solar chromosphere pose a number of challenges when we try to understand the coupling between the \emph{in situ} generation of various MHD wave modes in the photosphere, and their propagation into the overlying solar atmosphere. Since the chromosphere affects wave propagation and alters the overall plasma dynamics in an intrinsic manner, some of the main issues that need to be addressed may be outlined as follows:--\\
(i) Contemporary models of the solar chromosphere are still overly simple and do not take into account a number of important factors. Widely used one- or multi-fluid models need further improvement, particularly in the lower and upper chromospheric layers presenting  challenges for accurate models of the solar atmosphere. \\ 
Once this issue has been satisfactorily addressed, the next important task is to improve our knowledge and understanding of the drivers of MHD waves. Recent high-resolution ground-based observations (e.g., SST, ROSA, etc) reveal the presence of a variety of such drivers  (e.g., Bonet et al. 2008; Wedemeyer-B{\"o}hm et al. 2012; Morton et al. 2013; De Pontieu et al. 2014b; Liu et al. 2019). These drivers produce a variety of wave modes (kink, sausage, torsional modes) in  strongly magnetized flux tubes (see Section 3). Once MHD waves have been generated their propagation is influenced by the magnetic field and plasma properties of the chromospheric flux tube and the ambient medium. This can further enable the evolution of the mixed or individual modes (see sub-section 3.1) depending upon the local plasma conditions (e.g., Jess et al. 2009; McIntosh et al. 2011; Morton et al. 2012; Srivastava and Goossens 2013; Morton et al. 2015; Srivastava et al. 2017; Grant et al. 2018). Despite significant theoretical developments (Section 3), outstanding questions remain, specifically:\\
(ii) Which observables best reveal generation (e.g., resonant coupling of different tube modes in non-uniform media) and dissipation (e.g., resonant absorption) of MHD modes (e.g., kink/Alfv\'enic waves)? (sub-section 3.1, and, e.g., Goossens et al. 2011) \\
(iii) How can we determine the isolated symmetric modes of flux tubes (see sausage waves in sub-section 3.2) in the solar chromosphere when these are strongly structured in the radial direction?
(e.g., Lopin and Nagorny 2015; Chen et al. 2016; Li et al. 2018, 2020). \\
(iv) Are we currently in a position to estimate waves' exact roles in the heating/energy budget of the chromosphere, as well as that of the TR and inner corona?
Can we bring together observations and theory under realistic atmospheric conditions to provide a precise determination of wave energetics?\\
(v) How efficient are the various wave dissipation mechanisms in prominences and how important is the dissipated energy to the prominence energy balance? 


The answers to many of these outstanding questions will come through the continuous development and improvement of observational capabilities in the spatial, spectral and temporal and spatial domains. Spectropolarimetric observations from a wide range of  instruments (e.g., IRIS, 1m-SST, 4m-DKIST, upcoming 4m-EST and 2m-NLST) will allow us to constrain models of the chromosphere and its local dynamics in such a way that existing physical models will need to be adapted and refined, improving our understanding of the role of MHD waves.

The `whole chromosphere' perspective was emphasised in Section \ref{Sec:largescale}, where we saw that MHD waves in a magnetic atmosphere display a different nature as the relative importance of the sound and Alfv\'en speeds change drastically over 15 or more scale heights. The conversion of acoustic waves into magnetoacoustic waves in the lower chromosphere and their upward propagation crucially depends on the local plasma and magnetic field conditions, and the wave orientation with respect to the magnetic field. Some outstanding questions are:\\
(vi) Why are theoretical models unable to reproduce the observed variation of acoustic cutoff in the solar atmosphere? (e.g., Murawski et al. 2016; Wi{\'s}niewska et al. 2016; Felipe et al. 2018). \\
(vii) How do the transverse cut-off frequency and longitudinal cut-off frequency for longitudinal waves (magnetoacoustic waves, or sausage waves in flux tubes) and transverse waves (e.g., Alfv\'en waves; or Alfv\'enic/kink waves in tubes) respectively depend upon local plasma conditions (e.g., density, magnetic field and its inclination) and differ from the local acoustic cut-off frequency, and how do they further affect mode-coupling (McAteer et al. 2003)? 

Even as we continuously make progress refining our understanding on (vi) and (vii), new  imaging, spectral and spectropolarimetric observations challenge our models (e.g., Centeno et al. 2009; Heggland et al. 2011; Kayshap et al. 2018a, 2020). Further physical insight is yet required to broaden the entire scientific context of mass and energy supply in the solar atmosphere, which again poses a question: \\
(viii) Are omnipresent jets/plasma flows in the solar chromosphere responsible for plasma heating in addition to waves? How do these contribute to the fast component of the solar wind at higher altitudes? 

Despite significant observational and theoretical developments over two decades on magnetic waves, shocks, instabilities, and formation of the jets/flows, their respective roles or inter-relationships in a variety of chromospheric magnetic structures are not yet fully understood (e.g., Kudoh and Shibata 1999; De Pontieu et al. 2004; Shibata et al. 2007; Nishizuka et al. 2011; Srivastava et al. 2013; Iijima and Yokoyama 2015; Zhelyazkov et al. 2015; Brady and Arber 2016; Ku{\'z}ma et al. 2017; Kayshap et al. 2018b; Mishra and Srivastava 2019; W{\'o}jcik et al. 2019; Wang and Yokoyama 2020; Zaqarashvili et al. 2020). \\
(ix) The question of how the solar chromosphere is heated remains open. Do compressible (magnetoacoustic) waves make a significant contribution in this process in the large-scale chromosphere? Is the amount of thermal energy released during ion-neutral-electron collisions sufficient to balance the radiative losses? \\
(x) How does the thermal energy released by wave dissipation and plasma instabilities compare with thermal energy associated with the reconnection of magnetic field lines on various scales?

Answers to these outstanding questions will only come through simultaneous high spatial resolution observations over multiple heights. Specific tasks are to estimate the thickness of resonant layers (e.g., following on from Jess et al. 2020); resolve magnetic surfaces (iso-frequency surfaces) in pure torsional motions; determine characteristic length scales in instabilities; etc (e.g., see some science cases planned with DKIST; Rast et al. 2020). Moreover, spectral or spectro-polarimetric observations of the localized chromosphere should be made in such a manner that we detect the ultra-high cadence (temporal) variations revealing the high-frequency part of the (magneto-) acoustic and Alfv\'enic wave modes as they evolve in particular magnetic structures (e.g., see some science cases planned with DKIST; Rast et al. (2020). Moreover, there should be an appropriate estimation of the various cut-off frequencies with respect to the local acoustic cut-off frequency throughout the chromosphere, which sensitively depends upon the accurate estimation of density and magnetic field there, and also requires the consideration of appropriate model atmospheres (e.g., Avrett and Loeser 2008). Moreover, the most sensitive measurements of chromospheric magnetic fields should be utilized in re-constructing localized magnetic tubes along which these waves may channel their energy (e.g., Lagg et al. 2017).

These sophisticated measurements of plasma properties, and magnetic fields will lead to convincing observations of various physical properties of MHD waves, shocks, and instabilities (see Sections 2-4) in the localized chromosphere along with dynamical plasma processes (e.g., waves and flows) to determine their role in its heating and mass transport. The upcoming/existing ultra-high resolution observatories (e.g., 1m~SST, 4m~DKIST, 4m~EST, 2m~NLST, GREGOR, etc) from the ground, and space (e.g., IRIS, ADITYA-L1/SUIT), which are equipped with cutting-edge spectrographs/spectro-polarimeters, magnetograms, imagers, etc., will be able to fill important gaps in our observations. They will provide appropriate measurements and detection of dynamical processes in the solar chromosphere (e.g., waves, shocks, and instabilities), which will further refine our theoretical understanding/numerical modelling to answer many unresolved questions related to this highly dynamic layer of the solar atmosphere.
\\

\

\acknowledgments

We thank both the reviewers for their comments that improved our manuscript. JLB acknowledges support from MINECO and FEDER funds through project AYA2017-85465-P. JLB acknowledges discussions within the team on ``The eruption of solar filaments and the associated mass and energy transport'', led by JC Vial and PF Chen, and thanks ISSI for their support. 
The work of TVZ was funded by the Austrian Science Fund (FWF, project P30695-N27).
DBJ wishes to thank Invest NI and Randox Laboratories Ltd for the award of a Research \& Development Grant (059RDEN-1), in addition to the UK Science and Technology Facilities Council (STFC) for the award of a Consolidated Grant (ST/T00021X/1). DBJ also wishes to acknowledge scientific discussions with the Waves in the Lower Solar Atmosphere (WaLSA; https://www.WaLSA.team) team, which is supported by the Research Council of Norway and the Royal Society (Hooke18b/SCTM).
This research was supported by the Research Council of Norway through its Centres of Excellence scheme, project number 262622, and through grants of computing time from the Programme for Supercomputing.
KM's work was done within the framework of the projects from the Polish Science Center (NCN) Grant Nos. 2017/25/B/ST9/00506 and 
2020/37/B/ST9/00184. AKS and MM acknowledge support from the UK-India Education and Research Initiative under grant agreement  UGC-UKIERI-2017/18-014-A2. In this review paper, the data were not used, nor created for this research.


\end{document}